\documentclass[twocolumn,showpacs,preprintnumbers,amsmath,amssymb,fleqn]{revtex4}

\usepackage{graphicx}
\usepackage{bm}
\usepackage{subfigure}

\begin{document}

\title{The nature of the tensor order in Cd$_2$Re$_2$O$_7$}
\author{S. Di Matteo}
\affiliation{D\'epartement Mat\'eriaux Nanosciences, Institut de Physique de Rennes UMR UR1-CNRS 6251, Universit\'e de Rennes 1, F-35042 Rennes Cedex, France}
\author{M. R. Norman}
\affiliation{Materials Science Division, Argonne National Laboratory, Argonne, IL  60439, USA}

\date{\today}

\begin{abstract}
The pyrochlore metal Cd$_2$Re$_2$O$_7$ has been recently investigated by second-harmonic generation (SHG) reflectivity.
In this paper, we develop a general formalism that allows for the identification of the relevant tensor components of the SHG from azimuthal scans.
We demonstrate that the secondary order parameter identified by SHG at the structural phase transition is the $x^2-y^2$ component of the axial toroidal quadrupole.  This differs from the $3z^2-r^2$ symmetry of the atomic displacements associated with the $I\overline{4}m2$ crystal structure that was previously thought to be its origin.
Within the same formalism, we suggest that the primary order parameter detected in the SHG experiment is the $3z^2-r^2$ component of the magnetic quadrupole. We discuss the general mechanism driving the phase transition in our proposed framework, and suggest experiments, particularly resonant X-ray scattering ones, that could clarify this issue.
\end{abstract}


\maketitle

\section{Introduction}

Transition metal oxides with the pyrochlore structure A$_2$B$_2$O$_7$ have been intensively studied in the past twenty years, mainly because of their highly-frustrated magnetic lattice that leads to a rich phase diagram.  In some pyrochlores, the magnetic degrees of freedom can remain liquid-like without any long-range order \cite{tsune}, in others, the ground state degeneracy can be lifted via a phase transition that either lowers \cite{tcherny} or not \cite{arima} the symmetry of the lattice. Among these materials, Cd$_2$Os$_2$O$_7$ is characterized by a metal-insulator transition \cite{sleight} at 227 K, with an ordering of magnetic degrees of freedom but no lattice distortion \cite{mandrus}. The order parameter driving this transition was identified recently by resonant elastic X-ray scattering (REXS) \cite{yamaura}. It is a ferro-ordering of magnetic octupoles that breaks time-reversal symmetry without leading to macroscopic magnetization and without altering the crystal-lattice symmetry. This is also known as `all in - all out' magnetic order, where spins on a given Os tetrahedron either point inwards or outwards of the center, resulting in novel physical properties \cite{arima}.

Interestingly, the pyrochlore obtained by replacing Os with Re, Cd$_2$Re$_2$O$_7$, is also characterized by a phase transition at about the same temperature (T$_{c1} \sim 200$ K), yet instead is characterized by a large drop in the resistivity when moving into the lower temperature phase. In this case, the phase transition is also characterized by a crystal-symmetry reduction from cubic $Fd\overline{3}m$ to tetragonal $I\overline{4}m2$ \cite{hiroi}. This, along with the failure to observe magnetic order by nuclear magnetic resonance (NMR) \cite{nmr,sakai,arai}, led to the transition being interpreted in terms of a soft phonon mode \cite{kendzi}, identified as the parity-odd doublet ($E_u$) of the octahedral point group $O_h$ \cite{sergienko}. In this framework, the drop in the magnetic susceptibility \cite{hiroi2} at T$_{c1}$ was interpreted as a consequence of a reduction in the electronic density of states \cite{sergienko2}. Yet, the atomic displacements involved in the transition are so small (less than 0.005 ${\rm \AA}$ for the Re-O distance, less than 0.008 ${\rm \AA}$ for the Re-Re distance \cite{huang,note1}) that other mechanisms, leading to the crystal-symmetry reduction as a secondary effect, should not be excluded a priori. Moreover, the close analogy with the case of Cd$_2$Os$_2$O$_7$ might also suggest that magnetic degrees of freedom could play a role here as well, even though dipolar order has not been observed.

In this context, the recent study of the phase transition of Cd$_2$Re$_2$O$_7$ by second-harmonic generation (SHG) \cite{harter} provides a clear indication that a primary order parameter (OP) different from the softening of the phonon mode is the driving element of the phase transition. 
The SHG results were interpreted in terms of a primary, time-reversal even, nematic order parameter \cite{fu} (of symmetry $T_{2u}$) inducing a parity-breaking lattice distortion as a secondary order (of symmetry $E_{u}$). In order to achieve the necessary coupling in Landau theory (where the secondary OP goes as the square of the primary OP, as indicated by the temperature dependence of the SHG signal), a third OP is needed, of $T_{1g}$ symmetry, whose physical interpretation was unclear. All the irreducible representations, or irreps, here refer to the octahedral group $O_h$ of the high-temperature phase. 

As we will show below, the OP of $E_u$ symmetry detected in both SHG experiments conducted on Cd$_2$Re$_2$O$_7$ \cite{harter,petersen} is not the $E_u$ OP proposed in \cite{sergienko}, but rather an axial toroidal quadrupole of $x^2-y^2$ symmetry (in the cubic $Fd\overline{3}m$ coordinates, with $z$ along the tetragonal $c$ axis). 
Concerning the primary OP, we find that a time-reversal even scenario is not the only possibility. In order to have a time-reversal even primary OP, Harter {\it et al.}~\cite{harter} had to suppose a symmetry lowering to the point group $\overline{4}$, as a $T_{2u}$ primary OP is not allowed in the space group $I\overline{4}m2$ characterizing Cd$_2$Re$_2$O$_7$ below T$_{c1}$. 
But this would imply the simultaneous breakdown of the mirror symmetry and the two-fold axes in the $ab$ plane, inconsistent with the findings of X-ray \cite{hiroi,huang} and neutron \cite{dalton} diffraction.   
Instead, such a symmetry lowering is not necessary for the case of time-reversal odd OPs. Many experimental techniques, X-ray diffraction included, are blind to magnetic multipolar OPs, so that a time-reversal odd multipolar primary OP could be compatible with the absence of a magnetic signature at the transition as measured by NMR \cite{nmr,sakai,arai}.
In that context, we have found that two magnetic space groups could explain the SHG data without requiring a further lowering of the crystal symmetry: $I\overline{4}m'2'$ and $I\overline{4}'m'2$, both characterized by the breaking of the mirror symmetry for magnetic OPs, a key feature that could explain the SHG data, as discussed in Section II.B.

In order to clarify the above statements and provide a general framework to describe the phase transition of Cd$_2$Re$_2$O$_7$, the present paper is organized in three main sections. Section II is devoted to the symmetry analysis of the polarization tensors in an SHG experiment. This analysis, usually not performed in the literature (i.e., just the symmetry analysis of the matter tensor $\chi_{ijk}$ is performed), allows us to deduce the quadrupole nature of the primary OP by symmetry considerations and the requirement of no measured signal in an SS geometry \cite{notaSS}. In turn, this allows a straightforward explanation of the azimuthal dependence of the SHG intensity. Then, if we impose the requirement of no further reduction of the point group symmetry at the transition (i.e., not to $\overline{4}$), this naturally leads to the identification of the primary OP as a magnetic quadrupole. Our polarization analysis leads, as a byproduct, to the prediction of significant changes in the azimuthal scan if the incidence angle of the laser beam is varied.

In Section III, we focus on the analysis of all allowed magnetic tensor components compatible with the known experimental constraints,
and we propose magnetic patterns for the pyrochlore structure allowed by the two magnetic groups. For a summary of the work in Sections II and III, see Table \ref{summary}.
\begin{table}[ht!]
	\caption{Summary of the work in Sections II and III.  For each of the scenarios, non-magnetic ($I\overline{4}$) and two magnetic
	($I\overline{4}m'2'$ and $I\overline{4}'m'2$), we list the order parameters (OP, secondary or primary) along with their symmetry ($\pm$ is the symmetry under time reversal and $u$/$g$ the parity under spatial inversion), and in parenthesis their interpretation (ATQ is an axial toroidal quadrupole, MQ is a magnetic
	quadrupole, and MO is a magnetic octupole).  For the second scenario, $I\overline{4}m'2'$, one also has a toroidal octupole ($T_{2u}^-$).}
	\centering
	\begin{ruledtabular}
		\begin{tabular}{cccc}
			OP & $I\overline{4}$ & $I\overline{4}m'2'$ & $I\overline{4}'m'2$  \\
			\colrule
			secondary, $u$ & $E_{u}^+$ (ATQ) &  $E_{u}^+$ (ATQ) & $E_{u}^+$ (ATQ) \\
			primary, $u$ & $T_{2u}^+$ (?) & $T_{2u}^-$ (MQ) & $E_{u}^-$  (MQ) \\
			primary, $g$ & $T_{1g}^+$ (?) & $T_{1g}^-$ (MO) & $A_{2g}^-$  (MO) \\
		\end{tabular}
	\end{ruledtabular}
	\label{summary}
\end{table} 

In Section IV, we evaluate, both analytically and numerically, the outcomes of REXS experiments that could provide an experimental verification of our model and confirm the existence of ferro-quadrupolar magnetic order. The numerical analysis is performed via the FDMNES program \cite{joly} with the proposed magnetic structures as input. As a byproduct of our analysis, we also address the existence (or not) of a second phase transition in Cd$_2$Re$_2$O$_7$ around T$_{c2} \sim 120$ K that has been claimed in the literature \cite{hiroi} but for which there is no clear consensus \cite{disorder}. We suggest a key REXS experiment to clarify this issue as well.
Finally, in Section V, we provide our conclusions.

\section{Analysis of SHG in Cd$_2$Re$_2$O$_7$}

Second-harmonic generation is a third-order process in the matter-radiation interaction, determined by two absorptions of a photon $\hbar\omega$ and the emission of a photon $2\hbar\omega$ \cite{fiebig}. Its total scattering amplitude, $A_{SHG}$, can be written in quantum-mechanical terms using third-order perturbation theory \cite{prbIr}, instead of the semiclassical approach usually adopted in the optics literature \cite{boyd}. The advantage of the quantum mechanical approach, compared to the semi-classical one, is twofold.  First, it makes ab initio quantum mechanical calculations possible. Second, and more important for the present work, it allows representing the SHG signal as a scalar coupling of tensors describing the properties of the material with the corresponding tensors describing the electromagnetic field (see Eq.~(\ref{shgspher2}) below), in full analogy with the REXS case \cite{jphysD}, where it proved extremely useful in identifying multipolar orders. In particular, as demonstrated below, such a coupling highlights a general property of SHG signals for purely electric dipole transitions (E1-E1-E1): in the SS channel, SHG is blind to quadrupolar OPs. So the absence of an SHG signal in the SS channel, as in the case of the SHG experiments for Cd$_2$Re$_2$O$_7$, necessarily points to a quadrupolar OP. It is remarkable that E1-E1-M1 and E1-E1-E2 transitions do not have this property (M1 is a magnetic-dipole transition, E2 an electric-quadrupole transition). Therefore the absence of an SS signal and the presence of an SP signal is a signature of E1-E1-E1 transitions.

\subsection{General derivation of the SHG signal}

The full cross-section for SHG and its explicit derivation have been reported in Appendix A.1 and in Section III of Di Matteo and Norman \cite{prbIr} (DMN in the following). Here we recall some of the results that can be useful for the present analysis, in particular the analogy with the REXS tensor interpretation \cite{jphysD}. We also explicitly write the coupling terms of the susceptibility tensor, $\chi_{ijk}$, for each spherical-tensor component, that was not done in DMN.
Given the absence of an inversion center in the low-temperature phase of Cd$_2$Re$_2$O$_7$, and because of the absence of an SS signal, we can neglect magnetic dipole and electric quadrupole SHG transitions and just focus on the E1-E1-E1 SHG amplitude, written as: 

\begin{align} \label{shgspher}
A_{SHG}^{(e)} \propto  \tilde{\chi}_{\alpha\beta\gamma} \epsilon^o_{\alpha}\epsilon^i_{\beta}\epsilon^i_{\gamma} =\frac{1}{2} (\tilde{\chi}_{\alpha\beta\gamma} +  \tilde{\chi}_{\alpha\gamma\beta} ) \epsilon^o_{\alpha}\epsilon^i_{\beta}\epsilon^i_{\gamma}   
\end{align}
where $\tilde{\chi}_{\alpha\beta\gamma}$ represents the susceptibility leading to the SHG field (here, $i$ and $o$ refer to the incoming and outgoing polarizations of the electromagnetic field). Its full expression is given in DMN. The second equality in Eq.~(\ref{shgspher}) is a consequence of the symmetry of the incoming polarization ($\epsilon^o_{\alpha}\epsilon^i_{\beta}\epsilon^i_{\gamma}=\epsilon^o_{\alpha}\epsilon^i_{\gamma}\epsilon^i_{\beta}$), and it implies that SHG experiments are only sensitive to the part of the $\tilde{\chi}_{\alpha\beta\gamma}$ tensor that is symmetrized in the last two indices. In what follows, we call it ${\chi}_{\alpha\beta\gamma}$ (so ${\chi}_{\alpha\beta\gamma}={\chi}_{\alpha\gamma\beta}$). As the overall amplitude $A_{SHG}$ is a scalar, Eq.~(\ref{shgspher}) can be written as a scalar product of the corresponding irreducible representations of the three-dimensional rotation group, SO(3). 
Following the results of Section A.3 in DMN, the involved irreducible representations of the SO(3) group are: two dipoles ($l=1$), a quadrupole ($l=2$) and an octupole ($l=3$). This can be obtained by first coupling the two identical vectors $\epsilon^i_{\beta}$ and $\epsilon^i_{\gamma}$: the antisymmetric (vector) part is zero and only the scalar and the symmetric rank 2 tensor appear. Their coupling with $\epsilon^o_{\alpha}$ leads to the above result.
Therefore, Eq.~(\ref{shgspher}) can be written as:

\begin{align} \label{shgspher2}
A_{SHG}^{(e)} \propto  \sum_{i=1}^2{\chi}_{l=1}^{(i)}\tilde{O}_{l=1}^{(i)} + {\chi}_{l=2}\tilde{O}_{l=2} + {\chi}_{l=3}\tilde{O}_{l=3} 
\end{align}

Given their importance in describing the azimuthal scans in Cd$_2$Re$_2$O$_7$, we list the expressions for the five quadrupole components ($\tilde{O}_{l=2}$ for the polarization, ${\chi}_{l=2}$ for the susceptibilities). Dipoles and octupoles can be found in Appendix A. Note Eq.~(\ref{shgspher2}) implies a sum over all components (for $l=2$, $3z^2-r^2$, $x^2-y^2$, $xy$, $xz$, $yz$). The polarization dependence and the corresponding linear combination of susceptibilities $\chi_{ijk}$ for the five quadrupole components are:

\begin{align} \label{magquadcart}
& \tilde{O}_{3z^2-r^2} =  \epsilon_z^i (\vec{\epsilon}^o \times \vec{\epsilon}^i)_z \leftrightarrow \chi_{3z^2-r^2} =  \nonumber \\
& (\chi_{xyz}+\chi_{xzy}-\chi_{yzx}-\chi_{yxz})/2 ; \nonumber \\
& \tilde{O}_{x^2-y^2}  =\frac{1}{\sqrt{3}} \big[ \epsilon_x^i (\vec{\epsilon}^o \times \vec{\epsilon}^i)_x - \epsilon_y^i (\vec{\epsilon}^o \times \vec{\epsilon}^i)_y \big] \leftrightarrow \chi_{x^2-y^2} = \nonumber \\
&  (\chi_{xyz}+\chi_{xzy}+\chi_{yzx}+\chi_{yxz}-2\chi_{zxy}-2\chi_{zyx})/(2\sqrt{3}) ;  \nonumber \\
& \tilde{O}_{xy}  =\frac{1}{\sqrt{3}} \big[ \epsilon_x^i (\vec{\epsilon}^o \times \vec{\epsilon}^i)_y + \epsilon_y^i (\vec{\epsilon}^o \times \vec{\epsilon}^i)_x \big] \leftrightarrow \chi_{xy} = \nonumber \\
&  (\chi_{yyz}+\chi_{yzy}-\chi_{xzx}-\chi_{xxz}+2\chi_{zxx}-2\chi_{zyy})/(2\sqrt{3})  ; \nonumber \\
& \tilde{O}_{xz}  =\frac{1}{\sqrt{3}} \big[ \epsilon_z^i (\vec{\epsilon}^o \times \vec{\epsilon}^i)_x + \epsilon_x^i (\vec{\epsilon}^o \times \vec{\epsilon}^i)_z \big] \leftrightarrow \chi_{xz} = \nonumber \\
&  (\chi_{xxy}+\chi_{xyx}-\chi_{zyz}-\chi_{zzy}+2\chi_{yzz}-2\chi_{yxx})/(2\sqrt{3}) ; \nonumber \\
\end{align}
\begin{align}
& \tilde{O}_{yz}  = \frac{1}{\sqrt{3}} \big[ \epsilon_z^i (\vec{\epsilon}^o \times \vec{\epsilon}^i)_y + \epsilon_y^i (\vec{\epsilon}^o \times \vec{\epsilon}^i)_z \big] \leftrightarrow \chi_{yz} = \nonumber \\
&  (\chi_{zzx}+\chi_{zxz}-\chi_{yxy}-\chi_{yyx}+2\chi_{xyy}-2\chi_{xzz})/(2\sqrt{3}). \nonumber
\end{align}
with the usual definition of the vector product: $(\vec{\epsilon}^o \times \vec{\epsilon}^i)_z = \epsilon_{x}^o \epsilon_y^i - \epsilon_{y}^o \epsilon_{x}^i$.

We remind that the susceptibility multipoles corresponding to each polarization multipole, e.g., $(\chi_{xyz}+\chi_{xzy}-\chi_{yzx}-\chi_{yxz})/2$ for the $3z^2-r^2$ component, can have both a magnetic and a non-magnetic origin (see DMN, Section III). In both cases, only inversion-odd multipoles can be detected by E1-E1-E1 SHG. For example, the polarization term $\tilde{O}$ in Eq.~(\ref{magquadcart}), with the symmetry of a quadrupole, can be coupled either to a magnetic quadrupole, time-reversal odd and inversion odd, or to a non-magnetic quadrupole, time-reversal even and inversion odd. Among the latter, we remind that the axial toroidal quadrupole characterizes X-ray natural circular dichroism (XNCD), and that a nematic quadrupole \cite{fu} has been proposed by Harter {\it et al.}~\cite{harter}. As anticipated above, a key feature of the quadrupole term is that it is blind to SS polarization, as clear from the vector product between incoming and outgoing polarizations that is highlighted in the formal expression of all five components of Eq.~(\ref{magquadcart}). This is a general property of SHG, independent of the point-group symmetries of the material to be studied (not yet specified in the above equations). This property is analogous, for example, to the well-known fact that, in REXS, magnetic dipoles cannot be detected in SS geometry, but only in the polarization rotated SP channel. We remark that, of all the terms of Eq.~(\ref{magquadcart}), the quadrupole is the only multipole that gives a zero SS signal (as clear from the direct inspection of dipole and octupole terms reported in Appendix A that have non-zero contributions in SS geometry, as well as for the E1-E1-M1 and E1-E1-E2 terms reported in DMN).

By reducing the symmetry from spherical to octahedral, the above irreps of the SO(3) group branch to the irreps of the $O_h$ group as follows: both dipoles become $T_{1u}$ irreps of $O_h$. The octupole branches to $T_{1u}$, $A_{2u}$ and $T_{2u}$ irreps. Finally, the quadrupole branches to a doublet, $E_u$, the first two terms of Eq.~(\ref{magquadcart}), and a triplet, $T_{2u}$, the last three terms of Eq.~(\ref{magquadcart}). 
We end with a technical remark that will be useful for the analysis of the quadrupole, Eq.~(\ref{magquadcart}). It is an axial spherical tensor of rank two, inversion odd. This implies that it has the opposite behavior under inversion than a polar spherical tensor of rank two, which is inversion even (we remind that inversion properties of polar spherical tensors go like $(-1)^p$, where $p$ is the rank of the tensor). This implies that if a polar spherical tensor of rank two (e.g., the electric quadrupole $Q$) has an invariant component $Q_{3z^2-r^2}$ under the $S_{4z}$ symmetry operation, for an axial tensor this will not be invariant any more, as $S_{4z}$ is the product of a rotation, $C_{4z}$ and inversion, $I$. Indeed, by direct inspection, $O_{3z^2-r^2}$ changes sign under $S_{4z}$, whereas $O_{x^2-y^2}$ is invariant, contrary to the general behavior of a polar spherical tensor of rank two ($Q_{3z^2-r^2}$ is invariant under $S_{4z}$ and $Q_{x^2-y^2}$ changes sign). This property will be used in the next subsection to classify the secondary OP in Cd$_2$Re$_2$O$_7$.

\subsection{Application to Cd$_2$Re$_2$O$_7$: the secondary OP as an axial toroidal quadrupole}

The considerations of the previous subsection are general, applicable to any point group. In this subsection, we shall apply them to the case of the $\overline{4}m2$ point group of the low-temperature phase of Cd$_2$Re$_2$O$_7$, as well as to some of its subgroups, in order to explain the SHG experiments. 
We first make an important notational remark: throughout this paper we label the SHG susceptibilities, $\chi_{xyz}$, with $x$, $y$ and $z$ referred to the cubic axes of $Fd\overline{3}m$, in keeping with previous papers \cite{harter,petersen}. This corresponds, in principle, to the $\overline{4}2m$ point group, not $\overline{4}m2$, the two being just a different description of the same geometry related by a $45^\circ$ rotation (see the end of Appendix B for further explanation). Yet, the space group of Cd$_2$Re$_2$O$_7$ below T$_{c1}$ is the tetragonal $I\overline{4}m2$ space-group (International Tables for Crystallography (ITC) No.~119), that is physically different from the tetragonal $I\overline{4}2m$ space-group (ITC No.~121). In what follows, with a slight abuse of notation, we shall omit this notational difference for the point groups and refer also to the point symmetry as $\overline{4}m2$ (the correct relation with the SHG susceptibilities in the tetragonal frame is given at the end of Appendix B). 

We remind that in the $\overline{4}m2$ point group, only the following susceptibilities are allowed (see Table II in Section III.A): $\chi_{xzy}=\chi_{xyz}$, $\chi_{yzx}=\chi_{yxz}$ and $\chi_{zxy}=\chi_{zyx}$. If we limit to a time-reversal even OP, also the constraint $\chi_{xzy}=\chi_{yzx}$ appears (see Section III.A). The SP spectrum measured by Harter {\it et al.}~\cite{harter} breaks the mirror symmetry of $\overline{4}m2$. This, in turn, calls for a second, independent, tensor component $\chi_{ijk}$ that breaks the mirror symmetry. Such a reduction can be achieved in two ways: either with a symmetry lowering to the point group $\overline{4}$ (the highest possible without the mirror plane), or by considering a magnetic OP associated with a time-reversed mirror symmetry. In the latter case, two magnetic groups are possible: $\overline{4}m'2'$ and $\overline{4}'m'2$. In both cases, the mirror symmetry is associated with time reversal so that any magnetic OP that is odd with respect to the mirror symmetry alone can in principle explain the symmetry breaking measured in the SP channel. In previous work \cite{harter}, the case of a non-magnetic primary OP was considered. In this paper, in the next section, we discuss instead the case of a magnetic primary OP.  However, we briefly recall the main features of the $\overline{4}$ scenario for future comparison.

Reducing the point symmetry to $\overline{4}$ allows the following susceptibility terms to appear: $\chi_{xxz}=\chi_{xzx}=-\chi_{yyz}=-\chi_{yzy}$ and $\chi_{zxx}=-\chi_{zyy}$. Only terms with one $z$ label are allowed, which makes null the quadrupole terms coupled to $\tilde{O}_{xz}$ and $\tilde{O}_{yz}$ in Eq.~(\ref{magquadcart}). The relations valid for time-reversal even OPs, $\chi_{xzy}=\chi_{yzx}$ and $\chi_{xyz}=\chi_{yxz}$, make null also the susceptibility that couples to $\tilde{O}_{3z^2-r^2}$ in Eq.~(\ref{magquadcart}). Extending the analysis to dipole and octupole terms as well (listed in Appendix A), a direct inspection shows that the only non-zero susceptibilities are those that couple to the quadrupole $\tilde{O}_{x^2-y^2}$ ($E_u$) and $\tilde{O}_{xy}$ ($T_{2u}$) or to the octupole $\tilde{O}_{xyz}$ ($A_{2u}$) and $\tilde{O}_{z(x^2-y^2)}$ ($T_{2u}$). Using the above relations, valid for the $\overline{4}$ point group, the matter tensors become: $2(\chi_{xyz}-\chi_{zxy})$ (for $\tilde{O}_{x^2-y^2}$); $2(\chi_{zxx}-\chi_{xxz})$ (for $\tilde{O}_{xy}$); $\chi_{zxy}+2\chi_{xyz}$ (for $\tilde{O}_{xyz}$); and $2\chi_{zxx}+4\chi_{xxz}$ (for $\tilde{O}_{z(x^2-y^2)}$). In this framework, the additional symmetries introduced in Ref.~\onlinecite{harter} in order not to have an SS signal, i.e., that $\chi_{zxy}=-2\chi_{xyz}$ and that $\chi_{zxx}=-2\chi_{xxz}$, can be understood as the requirement of an absence of the octupole OPs with symmetry $xyz$ and $z(x^2-y^2)$. As we shall see in Section III, this additional symmetry that needs to be imposed here will be automatically recovered with the magnetic group $\overline{4}'m'2$.

The explicit dependence of the azimuthal scans for all the non-zero tensors of the $\overline{4}$ point group is reported in Eq.~(\ref{azscanphi}). The same expressions will also be useful for the magnetic point groups $\overline{4}m'2'$ and $\overline{4}'m'2$ that we shall analyze in Section III.A. The detailed calculations are reported in Appendix B. We have, for all allowed SP and SS terms: 
\begin{align} \label{azscanphi}
& \tilde{O}_{3z^2-r^2}^{SP} =  -\frac{1}{3}\big(\sqrt{2}\cos\theta \sin\phi_H +\sin\theta \sin(2\phi_H)\big) \\
 & \tilde{O}_{x^2-y^2}^{SP}  = \frac{1}{3}\big(-\sqrt{2} \cos\theta \cos\phi_H +\sin\theta \cos(2\phi_H)\big) \nonumber \\
 & \tilde{O}_{xy}^{SP}  =  \frac{2}{3\sqrt{3}}\big(\frac{1}{\sqrt{2}}\cos\theta \sin\phi_H -\sin\theta \sin(2\phi_H)\big)  \nonumber \\
&  \tilde{O}_{z(x^2-y^2)}^{SP} =  \frac{1}{6\sqrt{3}}\big(\cos\theta (3 \sin(3\phi_H)-\sin\phi_H \big) \nonumber \\
& \hspace{.75in} +\sqrt{2}\sin\theta \sin(2\phi_H)) \nonumber \\
& \tilde{O}_{xyz}^{SP}  =  \frac{1}{3}\big(\cos\theta\cos(3\phi_H)-\frac{1}{\sqrt{2}}\sin\theta\big)   \nonumber \\
& \tilde{O}_{z(x^2-y^2)}^{SS}  = \frac{1}{2\sqrt{3}} \big(\cos(3\phi_H)-\cos\phi_H \big)  \nonumber \\
& \tilde{O}_{xyz}^{SS} = -\frac{1}{3} \sin(3\phi_H)  \nonumber 
\end{align}
Here $\phi_H$ is the azimuthal angle measured by Harter {\it et al.}~\cite{harter} (see Fig.~\ref{azhar} in Appendix B), whereas $\theta = 10^\circ$ is the incidence angle with respect to the surface normal used in the experiment. This value of $\theta$ makes $\cos\theta \simeq 0.98$, so that the $\cos\theta$ term is dominant compared to the $\sin\theta$ term, though the latter is not negligible ($\sin\theta \sim 0.17$). We remark that octupolar OPs, besides giving an SS signal, have the wrong azimuthal scan even in the SP channel, with a $\sin(3\phi_H)$ or $\cos(3\phi_H)$ behavior.
In order to recover the experimental azimuthal scan, two OPs are required \cite{harter}, one going like $\sin\phi_H$ and the other going like $\cos\phi_H$. The former is associated with the primary OP and the latter with the secondary OP. Eq.~(\ref{azscanphi}) shows that there is only one possibility for the secondary OP, that has necessarily the symmetry of an $x^2-y^2$ quadrupole. Instead, the primary OP could have either the symmetry of a $3z^2-r^2$ quadrupole or that of an $xy$ quadrupole, both going like $\sin\phi_H$. The latter belongs to a $T_{2u}$ irrep, compatible with the Landau free energy proposed in Harter {\it et al.}~\cite{harter}. The former does not contribute in the $\overline{4}$ point group as stated above (it couples to $\chi_{xyz}+\chi_{xzy}-\chi_{yzx}-\chi_{yxz}$ that is zero in $\overline{4}$ in the absence of magnetism). However, as we shall see in Section III.A, the inclusion of magnetic point groups makes the linear combination $\chi_{xyz}+\chi_{xzy}-\chi_{yzx}-\chi_{yxz}$ associated with the $3z^2-r^2$ quadrupole non-zero and its symmetry ($E_u$) is allowed in the free energy.
We shall discuss about the magnetic origin of the primary OP in Section III. Here, we discuss the physical interpretation of the secondary OP, an $x^2-y^2$ axial toroidal quadrupole. We remark that it is not the same OP as the atomic displacements of $3z^2-r^2$ symmetry that have been proposed before to explain the SHG secondary OP \cite{petersen,harter}.

In order to understand the physical origin of the SHG signal determined by the secondary OP, consider the following model.
We start with the positions of Re ions in the cubic $Fd\overline{3}m$ phase \cite{sergienkoJPSJ}:

\begin{align} \label{positions}
& \vec{r}_1=(1/8,1/8,1/8) &  \vec{r}_2=(-1/8,-1/8,1/8) \\
& \vec{r}_3=(-1/8,1/8,-1/8) &  \vec{r}_4=(1/8,-1/8,-1/8) \nonumber
\end{align}

We have also the following displacements from the cubic positions in the $I\overline{4}m2$ and $I4_122$ phases \cite{sergienkoJPSJ}:

\begin{align} \label{displs}
&\vec{\delta}_1=(x_1,y_1,z_1) & \vec{\delta}_2=(-x_1,-y_1,z_1) \\
&\vec{\delta}_3=(-x_1,y_1,-z_1) & \vec{\delta}_4=(x_1,-y_1,-z_1) \nonumber
\end{align}
We remind that for $I\overline{4}m2$, we have the relation $x_1=y_1$ (and $x_1$ and $z_1$ have opposite signs), and for $I4_122$ we have $x_1=-y_1$ and $z_1=0$. Though $I4_122$ symmetry has been excluded at $T_{c1}$, it is useful to consider it for further analysis in Section IV.

We can now evaluate the axial toroidal dipole \cite{dubovik} determined by these displacements for the tetrahedral cell. We remark that the value of this vector is independent of the choice of the origin, because the tetrahedron is non-polar even after the displacements (i.e., $\sum_{i=1}^4 \vec{\delta}_i=0$). The values we get for each tetrahedron are equal to one-half those of the unit cell in the tetragonal phase, because of the bcc translation.
Using $\vec{g}=\sum_{i=1}^4 \vec{g}_i$, with $\vec{g}_i=\vec{r}_i \times \vec{\delta}_i$ we obtain:
\begin{align} \label{tordip}
& \vec{g}_1=\frac{1}{8}(z_1-y_1,x_1-z_1,y_1-x_1) \\
& \vec{g}_2=\frac{1}{8}(-z_1+y_1,-x_1+z_1,y_1-x_1) \nonumber \\
& \vec{g}_3=\frac{1}{8}(-z_1+y_1,x_1-z_1,-y_1+x_1) \nonumber \\
& \vec{g}_4=\frac{1}{8}(z_1-y_1,-x_1+z_1,-y_1+x_1) \nonumber
\end{align}
which gives, for the total axial toroidal dipole of the tetrahedron $\vec{g}=\sum_{i=1}^4 \vec{g}_i=0$, as expected for both $\overline{4}m2$ and $422$ symmetries. We can now evaluate the corresponding axial toroidal quadrupoles, $G_{ij}$:
\begin{align} \label{torquad}
& G_{xy}\equiv \frac{1}{4} \sum_{i=1}^4 r_{ix}g_{iy}+r_{iy}g_{ix} = 0 \\
& G_{xz}\equiv \frac{1}{4} \sum_{i=1}^4 r_{ix}g_{iz}+r_{iz}g_{ix} = 0 \\
& G_{yz}\equiv \frac{1}{4} \sum_{i=1}^4 r_{iz}g_{iy}+r_{iy}g_{iz} = 0  \\
& G_{3z^2-r^2}\equiv \frac{1}{4} \sum_{i=1}^4 3r_{iz}g_{iz}-\vec{r}\cdot\vec{g} \propto x_1 - y_1 \\
& G_{x^2-y^2}\equiv \frac{1}{4} \sum_{i=1}^4 r_{ix}g_{ix}-r_{iy}g_{iy} \propto 2z_1 - x_1 - y_1  
\end{align}

\begin{figure}[ht!]
\includegraphics[width=0.4\textwidth]{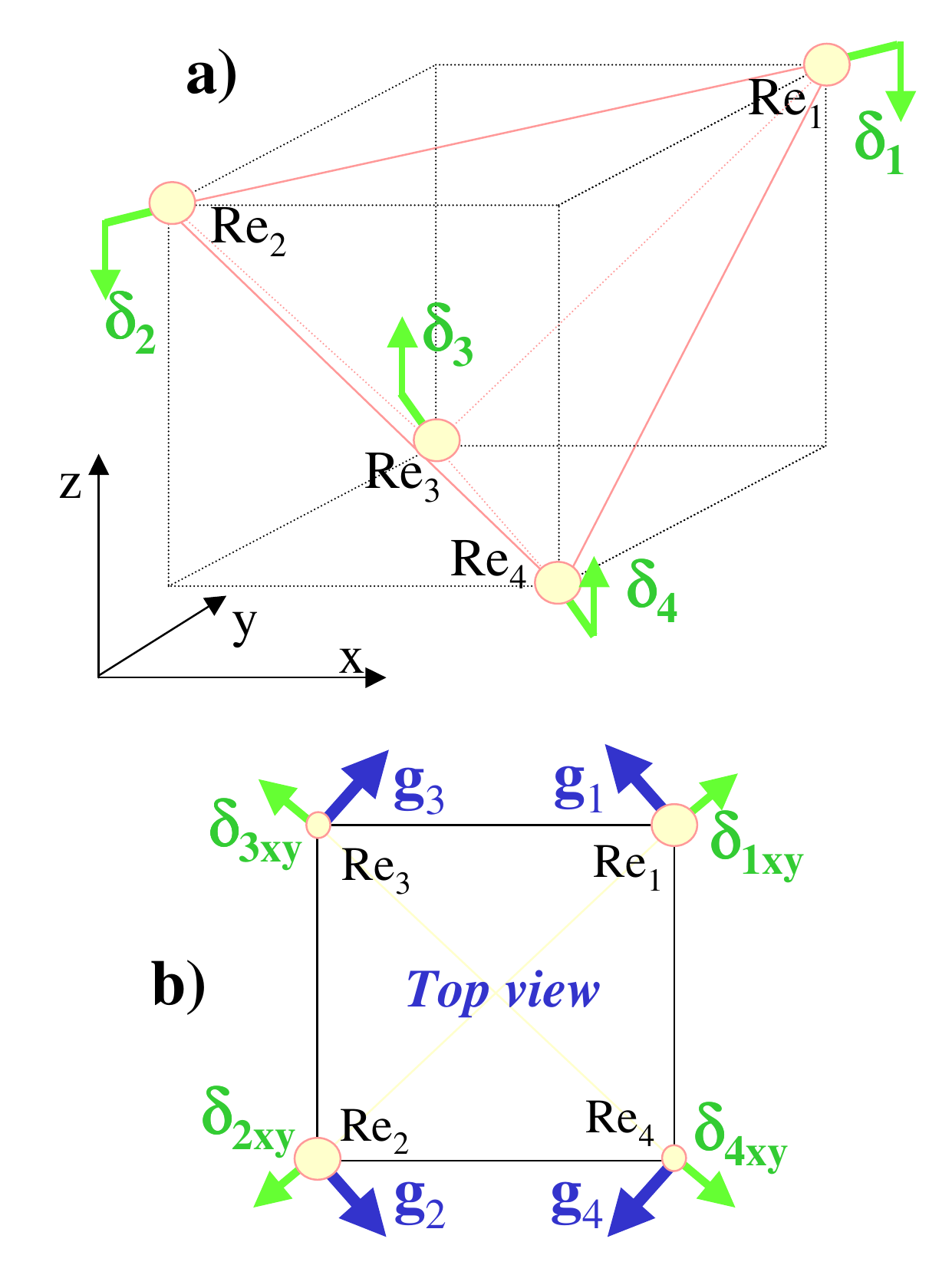}
\caption {(a) The unit tetrahedron with the Re$_i$ positions of Eq.~(\ref{positions}) and the displacements of Eq.~(\ref{displs}), depicted as green arrows. (b) View from the top of the tetrahedron, with the axial toroidal dipoles of Eq.~(\ref{tordip}) at each Re$_i$ position depicted as blue arrows. Both planes (z=$\pm\frac{1}{8}$) contribute one-half to the total axial toroidal quadrupole.}
\label{axtor}
\end{figure}

Notice that $G_{x^2-y^2}\neq 0$ and $G_{3z^2-r^2}=0$ in $I\overline{4}m2$ because of the condition $x_1=y_1$ and the opposite sign of $z_1$ and $x_1$. In the other case ($I4_122$), we would have had instead $G_{3z^2-r^2}\neq 0$ and $G_{x^2-y^2}=0$, because of the condition $x_1=-y_1$ and $z_1=0$.

Now we can solve the apparent contradiction of ${3z^2-r^2}$ versus $x^2-y^2$ OPs: the coordinate dependence of the axial toroidal quadrupole $G_{x^2-y^2}\propto x_1 + y_1 -  2z_1$ is the same as that of the $E_u$ ($3z^2-r^2$) component of the Re atomic displacements \cite{sergienko} associated with $I\overline{4}m2$. Here, though, $G$ is a quadrupole of $x^2-y^2$ symmetry and for this reason is associated with the polarization spherical tensor $\tilde{O}_{x^2-y^2}$. The reason for this change in the nature of the tensor detected by SHG ($x^2-y^2$ instead of $3z^2-r^2$) was already sketched at the end of Section II.A. The displacement OP \cite{sergienko} is a polar spherical tensor (the displacement is a polar vector) whose behavior under the symmetry operations of $I\overline{4}m2$ are such that, in particular, $3z^2-r^2$ is invariant under the $S_{4z}$ symmetry operation, whereas $x^2-y^2$ changes sign. So, as correctly found in the literature \cite{sergienko}, the transition to the $I\overline{4}m2$ space group is characterized by a $3z^2-r^2$ polar tensor, as the $x^2-y^2$ polar tensor would not be invariant.
However, SHG is not sensitive to a rank-two spherical polar tensor, but to a rank-two spherical axial tensor, that we have identified as the axial toroidal quadrupole. Under the $S_{4z}\equiv I C_{4z}$ symmetry operation, the inversion symmetry behavior is opposite for polar and axial tensors. Therefore, for an axial rank-two tensor, $x^2-y^2$ is invariant under the $S_{4z}$ symmetry operation, whereas $3z^2-r^2$ changes sign, in keeping with the above discussion.
We remark that our findings are further confirmed by the fact that $I4_122$ allows for an XNCD signal, whereas $I\overline{4}m2$ does not \cite{mike}, since as is well known \cite{sergioxncd}, when the X-ray beam is along the tetragonal $c$-axis, XNCD is sensitive to the $G_{3z^2-r^2}$ component of the axial toroidal quadrupole and not to $G_{x^2-y^2}$. A pictorial representation of the axial toroidal quadrupole for the single tetrahedron is reproduced in Fig.~\ref{axtor}.

\begin{figure}[ht!]
\includegraphics[width=0.45\textwidth]{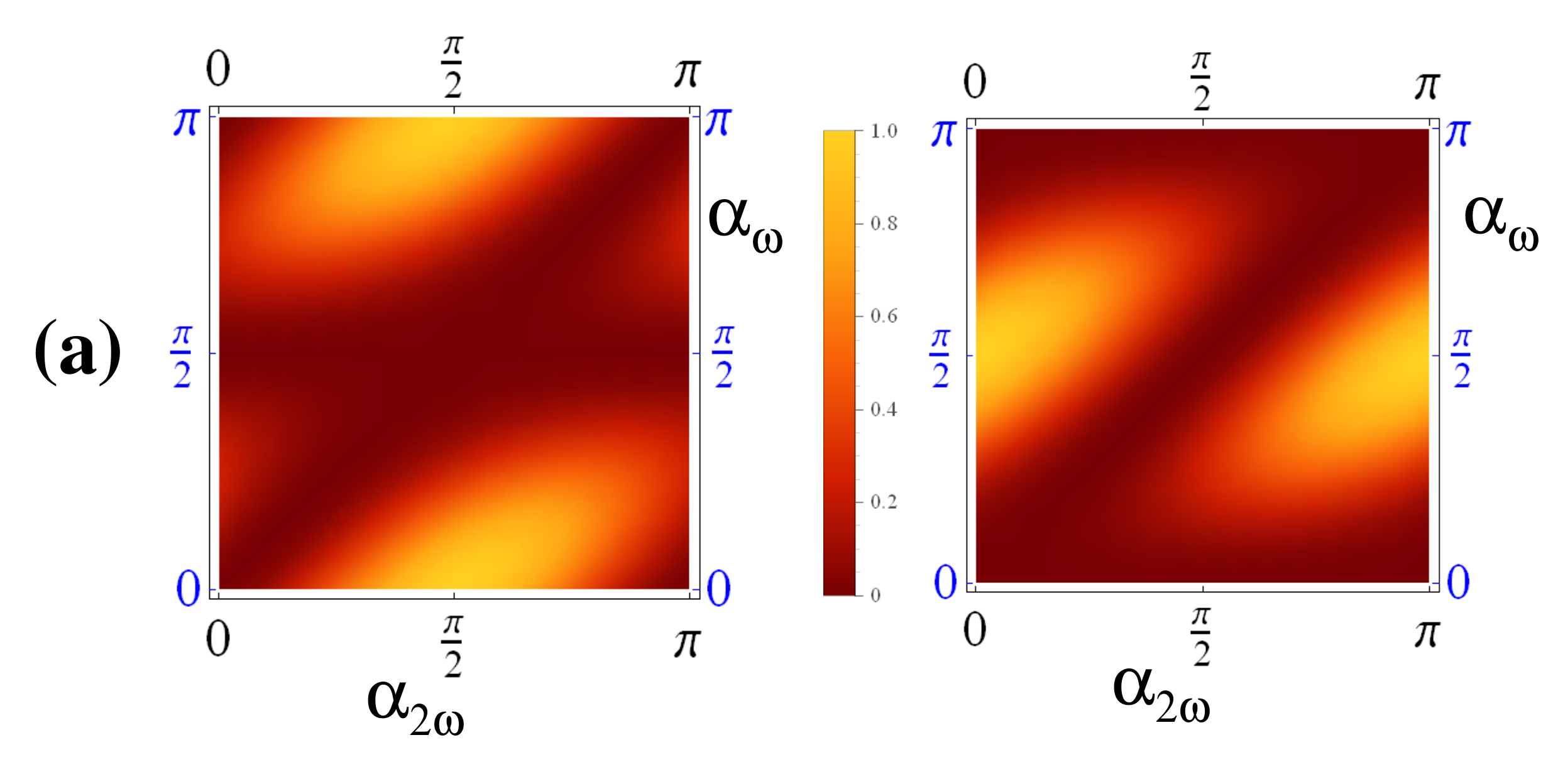}
\includegraphics[width=0.45\textwidth]{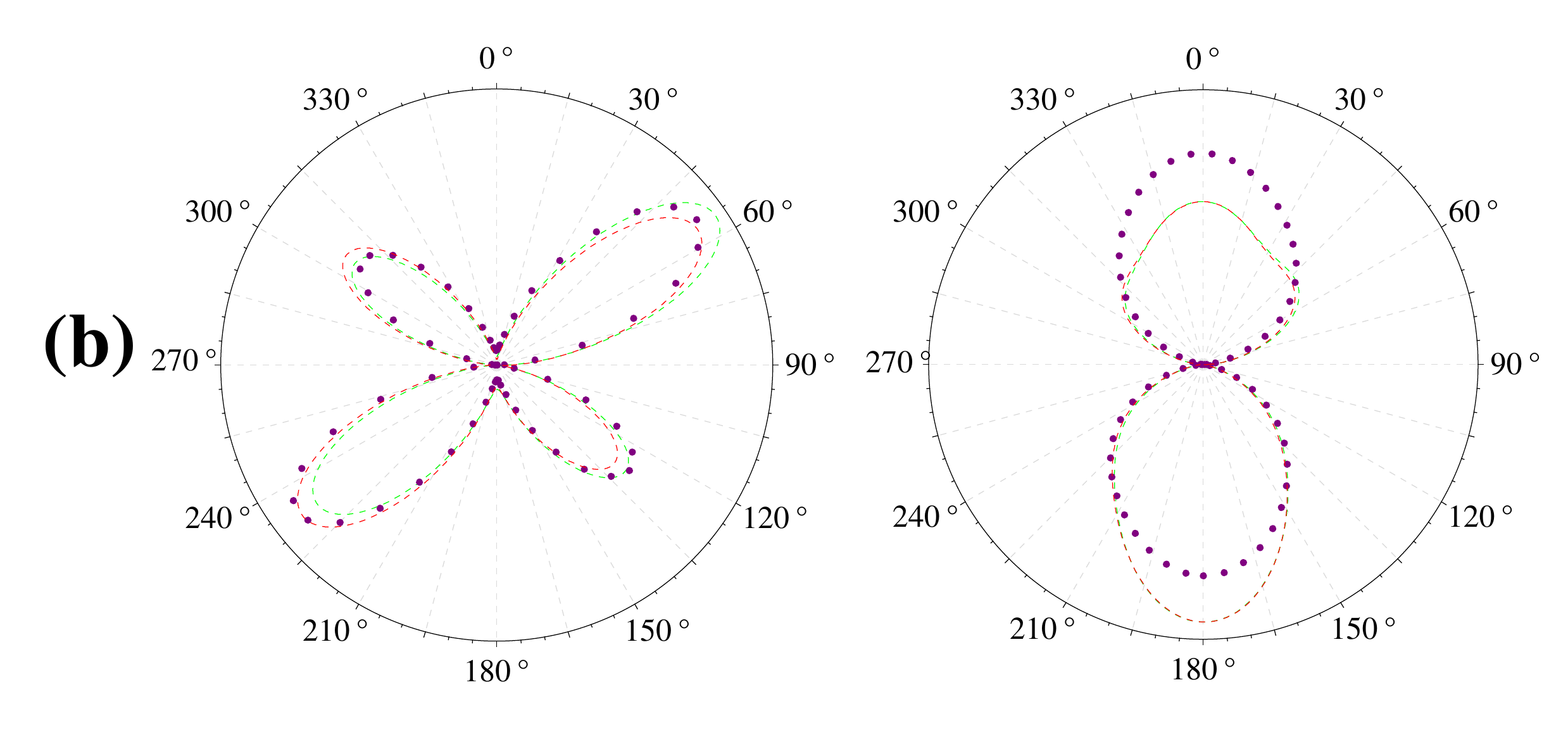}
\caption {(a) Left: description of Petersen {\it et al.}~SHG experiment \cite{petersen} from $\tilde{O}_{x^2-y^2}^{SP}$ of Eq.~(\ref{azscanphi}) (without the birefringence term, see Appendix B). Right: The result if $\tilde{O}_{3z^2-r^2}^{SP}$, that would be active in the $I4_122$ space group, was used instead. (b) Dots represent the fit of Harter {\it et al.}~\cite{harter} to their data at T=199.7 K (left) and at 196.6 K (right). The azimuthal angle is $\phi_H$ (see Fig.~\ref{azhar}). Curves are our fit, for $\theta=10^\circ$, from term $\tilde{O}_{x^2-y^2}^{SP}$ for the secondary OP and term $\tilde{O}_{3z^2-r^2}^{SP}$ (dashed green) {\it or} $\tilde{O}_{xy}^{SP}$ (dashed red) for the primary OP.}
\label{figpet}
\end{figure}

\begin{figure}[ht!]
\includegraphics[width=0.45\textwidth]{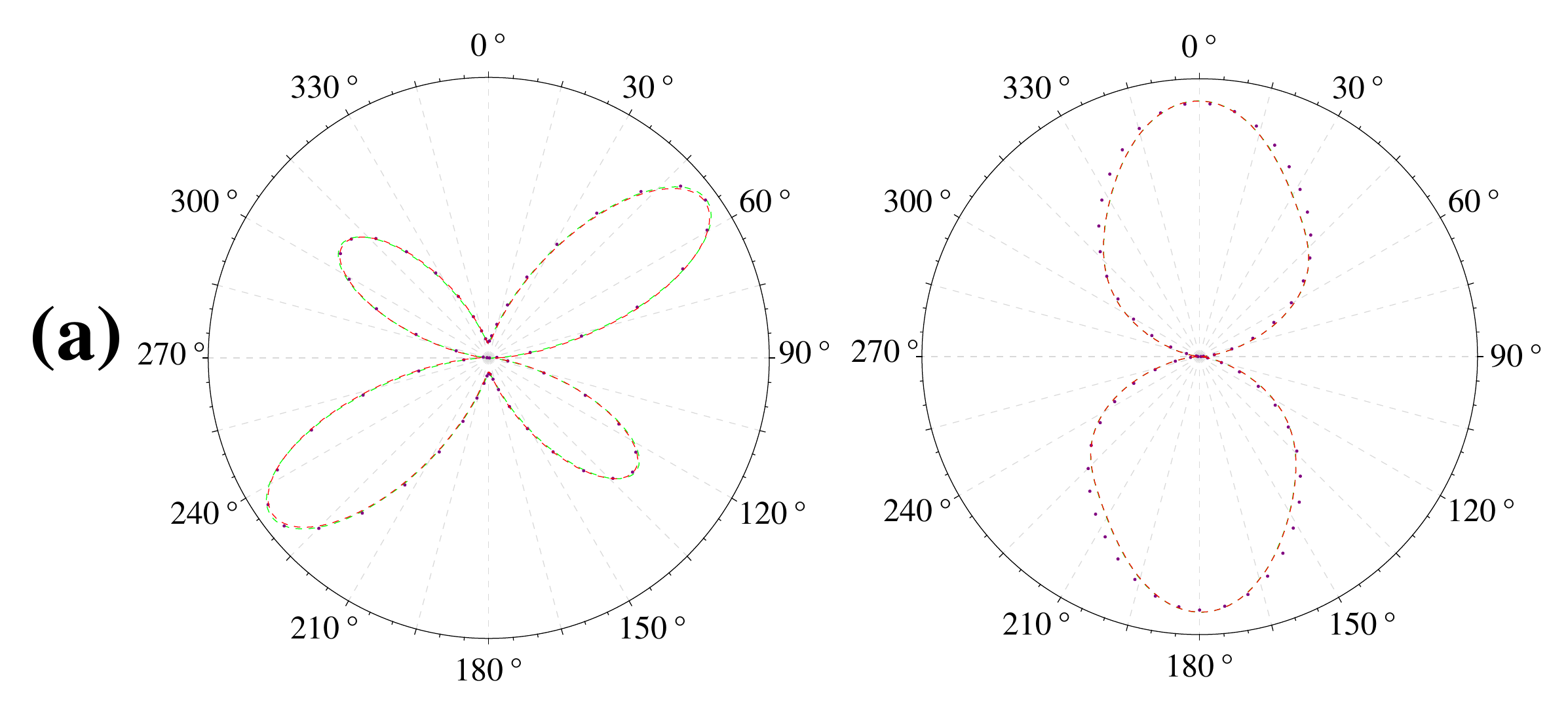}
\includegraphics[width=0.45\textwidth]{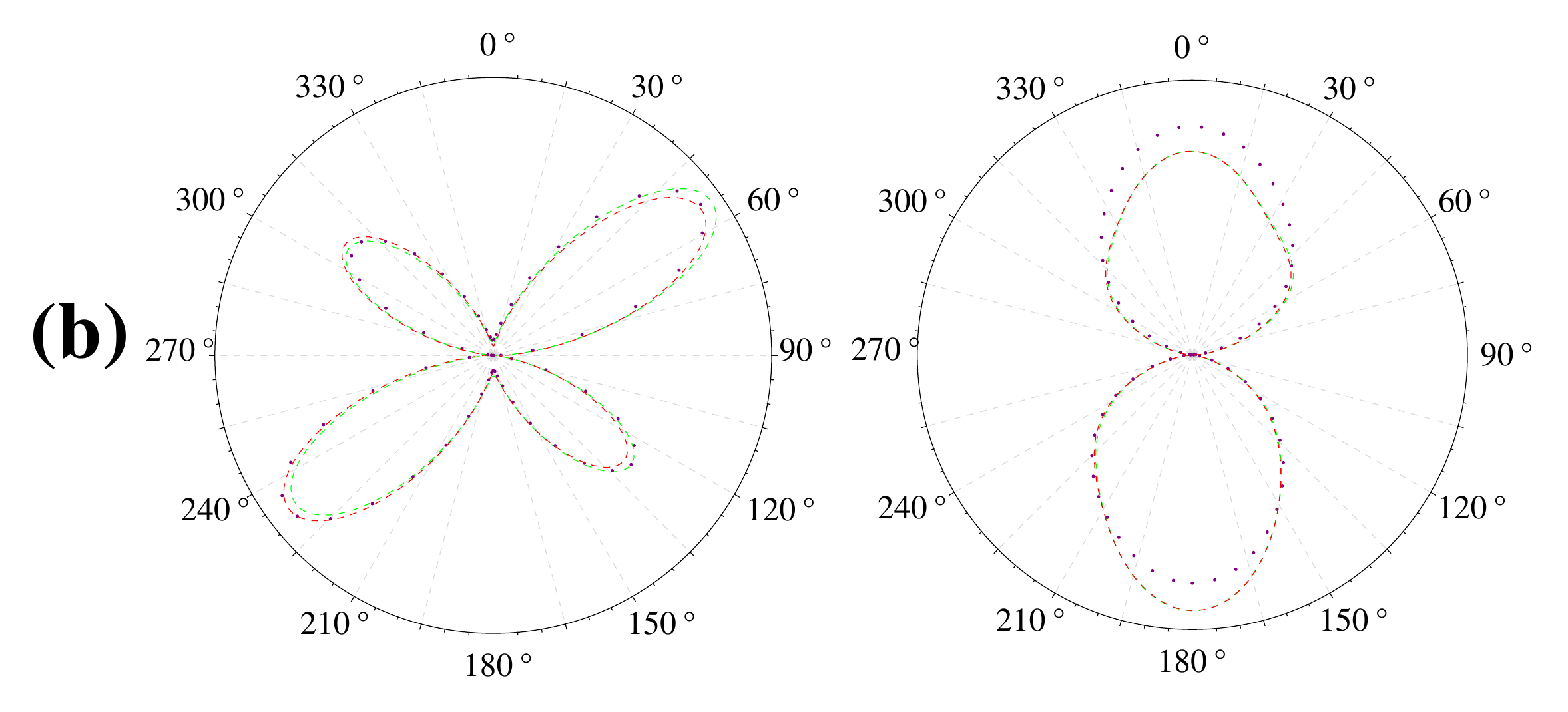}
\includegraphics[width=0.45\textwidth]{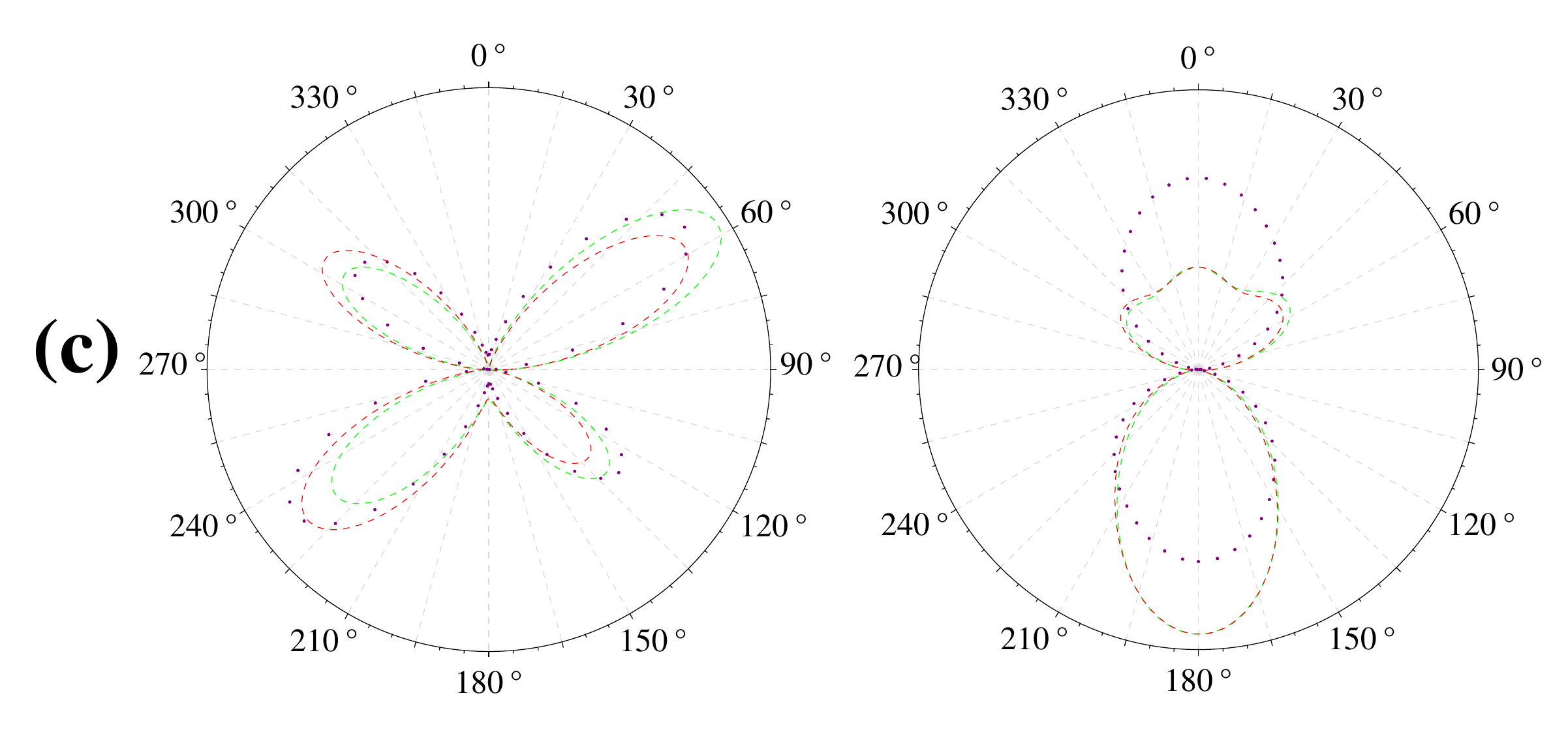}
\includegraphics[width=0.45\textwidth]{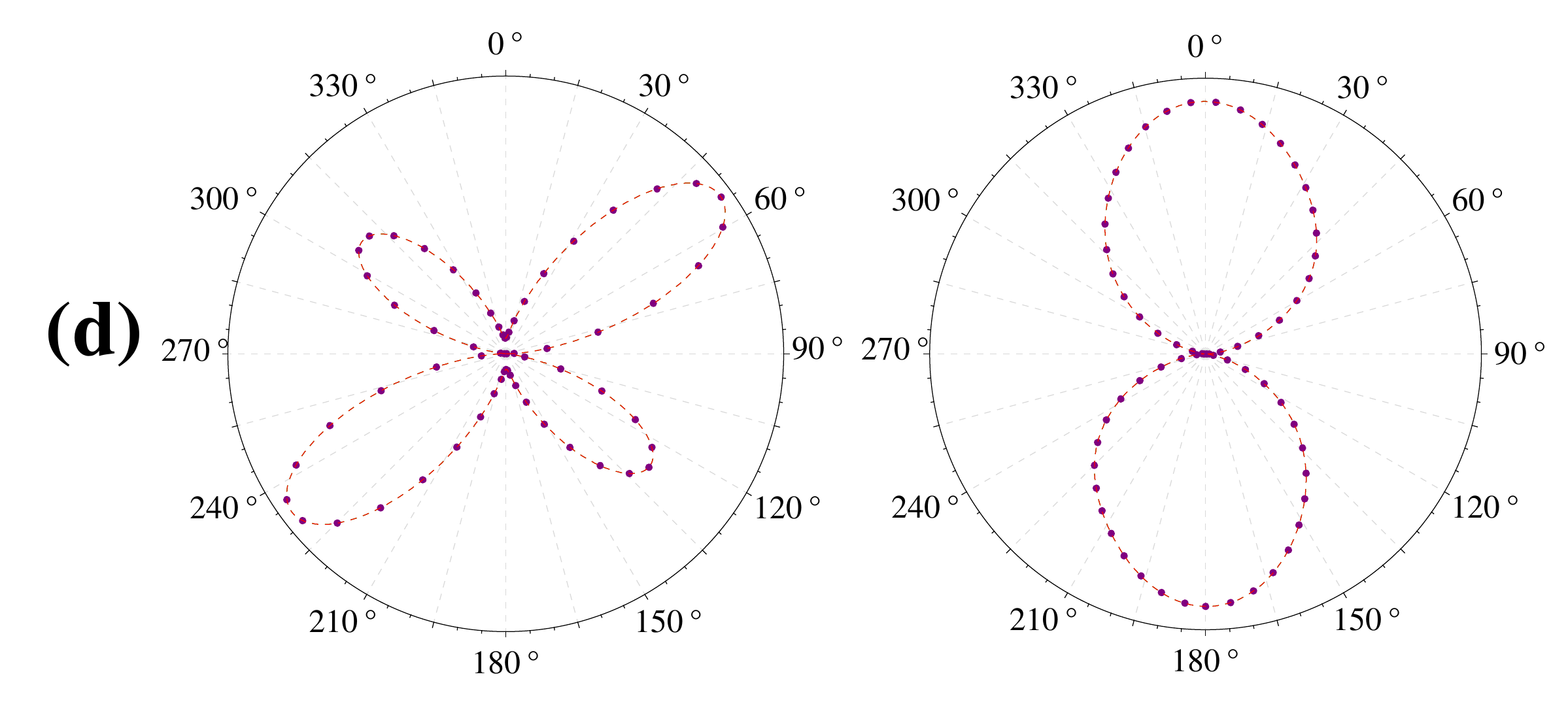}
\caption {Same as Fig.~\ref{figpet}b, but for (a) $\theta=0^\circ$, (b) $\theta=5^\circ$, and (c) $\theta=20^\circ$.  For these three plots, the fit coefficients are the same as Fig.~\ref{figpet}b.  (d) A fit to the data where it is assumed that $\theta=0^\circ$.}
\label{fignew}
\end{figure}

We conclude this subsection by showing that both SHG experiments on Cd$_2$Re$_2$O$_7$ \cite{harter,petersen} are associated with the axial toroidal quadrupole calculated above.
The first SHG experiment, by Petersen {\it et al.}~\cite{petersen}, can be reproduced with the formula $\cos^2\alpha_{\omega} \sin^2(\alpha_{\omega}-\alpha_{2\omega})$, plotted in the left frame of Fig.~\ref{figpet}(a). The details of the calculation, valid for the the $I\overline{4}m2$ space group, and the experimental setup are reported in Appendix B. The whole signal is provided only by the $\tilde{O}_{x^2-y^2}^{SP}$ term, the second line of Eq.~(\ref{azscanphi}). It is therefore the same OP identified as the secondary OP in Harter {\it et al.}~\cite{harter} and it is associated with the axial toroidal quadrupole. We only remark that in the present approach, we have not considered the corrections reported in Ref.~\onlinecite{petersen} due to linear birefringence (see below and also the discussion in Appendix B).

Harter {\it et al.}'s azimuthal scan can be recovered as well by means of Eq.~(\ref{azscanphi}). 
Technical details are provided in Appendix B. We remark however that a clear discrepancy with the experimental data can be seen in the right panel of Fig.~\ref{figpet}(b), around $\phi=0$ and $\phi=180^\circ$. This is mainly determined by the interference of the $\cos\phi_H$ and $\cos(2\phi_H)$ functions in $\tilde{O}_{x^2-y^2}^{SP}$.
We can suggest two reasons for this discrepancy. First, the effective incident angle $\theta$ could be reduced with respect to its nominal value by refraction at the sample surface. If the effective $\theta$ is sufficiently small, Eq.~(\ref{azscanphi}) would properly describe the experimental data (Fig.~\ref{fignew}). 
Second, the interference can be reduced if one component of the outgoing beam is dephased, say due to linear birefringence \cite{petersen}. If, for example, $i$ multiplied the third component ($\epsilon^{\rm out}_{z'}$) of the last line of Eq.~(\ref{polax}), this dephasing by $\pi/2$ would propagate to Eq.~(\ref{azscanphi}), leading to a sum in quadrature of the $\cos\phi_H$ and $\cos(2\phi_H)$ terms (for the secondary OP) and of the $\sin\phi_H$ and $\sin(2\phi_H)$ terms (for the primary OP), leading to the azimuthal dependence measured by Harter {\it et al.}

The physical origin of the discrepancy (whether refraction or birefringence) might be established through Eq.~(\ref{azscanphi}), by performing the same experiment for larger values of $\theta$. If the origin of the effect were refraction, increasing $\theta$ would make the interference terms more effective and a strongly asymmetric pattern would appear (Fig.~\ref{fignew}). Interestingly, varying the incident angle $\theta$ also allows differentiating the two terms $\tilde{O}_{3z^2-r^2}^{SP}$ and $\tilde{O}_{xy}^{SP}$, reproduced in Fig.~\ref{figpet}(b), in dashed green and dashed red, respectively, and barely distinguishable for $\theta=10^\circ$. The two cases can instead be identified for $\theta=20^\circ$ from the left panel of Fig.~\ref{fignew}(c), as the $\tilde{O}_{3z^2-r^2}^{SP}$ term (dashed green) is symmetric in the small lobes and asymmetric in the big lobes, whereas the $\tilde{O}_{xy}^{SP}$ (dashed red) is symmetric in the big lobes and asymmetric in the small lobes.
The importance of the latter point is that, as demonstrated in the next section, the two terms are associated with two different magnetic groups and OPs. 
We finally remark that this difference would not appear if the discrepancy with the experimental data in Fig.~\ref{figpet}(b) was due to birefringence.

\section{Possible magnetic order for Cd$_2$Re$_2$O$_7$}

In this section, we analyze the magnetic subgroups of $I\overline{4}m2$ in order to determine a possible magnetic origin of the primary OP. 

\subsection{Magnetic point group of Cd$_2$Re$_2$O$_7$ below T$_{c1}$}

A possible magnetic origin of the quadrupole primary OP cannot be determined by the polarization analysis, since it produces the same azimuthal scan for time-reversal even and time-reversal odd quadrupolar OPs. As seen above, only two terms can play the role of primary OPs, those associated with the polarization terms $\tilde{O}_{xy}$ and $\tilde{O}_{3z^2-r^2}$.
Interestingly, there are two irreps of $\overline{4}m2$ that break the mirror symmetry, $A_{2}$ and $B_{1}$, the former with symmetry $\tilde{O}_{xy}$ and the latter with symmetry $\tilde{O}_{3z^2-r^2}$. Their character table is shown in Table \ref{chtable} below:

\begin{table}[ht!]
	\caption{Character table of $A_{2}$ and $B_{1}$ irreps of $\overline{4}m2$, corresponding to the $\tilde{O}_{xy}$ and $\tilde{O}_{3z^2-r^2}$ components of the quadrupole. For magnetic OPs, they become the totally symmetric irrep $A_{1}$ for $\overline{4}m'2'$ and $\overline{4}'m'2$, respectively.}
	\centering
	\begin{ruledtabular}
		\begin{tabular}{c|c|c|c|c|c}
			$\overline{4}m2$ & $E$ & $2S_4$ & $C_{2z}$ & $2C_{2x}$ & $2\sigma_d$  \\
			\colrule
			$A_{2}$ & $+1$ & $+1$ & $+1$ & $-1$ & $-1$  \\
						\colrule
			$B_{1}$ & $+1$ & $-1$ & $+1$ & $+1$ & $-1$  \\
		\end{tabular}
	\end{ruledtabular}
	\label{chtable}
\end{table} 

They become the totally symmetric irrep (i.e., $A_1$) for the magnetic groups $\overline{4}m'2'$ and $\overline{4}'m'2$, respectively. In fact, for magnetic OPs, the $\overline{4}m'2'$ point group changes the sign of $C_{2x}$ and $\sigma_d$ because of the associated time-reversal operation, making them positive and changing therefore $A_2$ into $A_1$. Analogously, the $\overline{4}'m'2$ point group changes the sign of $S_4$ and $\sigma_d$ for magnetic OPs because of the associated time-reversal operation, making them positive and changing therefore $B_1$ into $A_1$.

We remark that, in the case of reduction to the $\overline{4}$ point group \cite{harter}, the two symmetry operations $C_{2x}$ and $\sigma_d$ disappear, and with them also the two associated negative characters. So, in the $\overline{4}$ point group, the $A_2$ irrep of the susceptibilities associated with $\tilde{O}_{xy}$ becomes totally symmetric and allowed as an OP (but not $B_1$).  
 
We analyze now how the susceptibilities transform in the two magnetic groups $\overline{4}m'2'$ and $\overline{4}'m'2$. We add the parent group $\overline{4}m2$ for comparison. For simplicity, we remove the identity and consider only the terms with one $z$ label, whose azimuthal scan is given by Eq.~(\ref{azscanphi}): $\chi_{xyz}$ and $\chi_{yxz}$ (allowed in $\overline{4}m2$), $\chi_{xxz}$ and $\chi_{yyz}$. 
For the non-magnetic group $\overline{4}m2$, we have Table \ref{susc1}.

\begin{table}[ht!]
	\caption{Transformation properties of the relevant susceptibilities for $\overline{4}m2$.}
	\centering
	\begin{ruledtabular}
		\begin{tabular}{c|c|c|c|c}
			$\overline{4}m2$ & $2S_4$ & $C_{2z}$ & $2C_{2x}$ & $2\sigma_d$  \\
			\colrule
			$\chi_{xyz}$ & $\chi_{yxz}$ & $\chi_{xyz}$ & $\chi_{xyz}$ & $\chi_{yxz}$  \\
						\colrule
			$\chi_{yxz}$ & $\chi_{xyz}$ & $\chi_{yxz}$ & $\chi_{yxz}$ & $\chi_{xyz}$  \\
			\colrule
			$\chi_{xxz}$ & $-\chi_{yyz}$ & $\chi_{xxz}$ & $-\chi_{xxz}$ & $\chi_{yyz}$  \\
						\colrule
			$\chi_{yyz}$ & $-\chi_{xxz}$ & $\chi_{yyz}$ & $-\chi_{yyz}$ & $\chi_{xxz}$  \\
		\end{tabular}
	\end{ruledtabular}
	\label{susc1}
\end{table} 

From the first two components of Table \ref{susc1}, we obtain that the linear combination $\chi_{xyz}+\chi_{yxz}$ behaves like the totally symmetric irrep, whereas $\chi_{xyz}-\chi_{yxz}=0$ (already used in the previous section). Similarly, from the last two components, we get $\chi_{xxz}=-\chi_{xxz}=0$ and $\chi_{yyz}=-\chi_{yyz}=0$, as already known. Doing this for the magnetic subgroups, we have instead Tables \ref{susc2} and \ref{susc3}.

 \begin{table}[ht!]
	\caption{Transformation properties of the relevant susceptibilities for $\overline{4}m'2'$.}
	\centering
	\begin{ruledtabular}
		\begin{tabular}{c|c|c|c|c}
			$\overline{4}m'2'$ & $2S_4$ & $C_{2z}$ & $2TC_{2x}$ & $2T\sigma_d$  \\
			\colrule
			$\chi_{xyz}$ & $\chi_{yxz}$ & $\chi_{xyz}$ & $\chi_{xyz}^*$ & $\chi_{yxz}^*$  \\
						\colrule
			$\chi_{yxz}$ & $\chi_{xyz}$ & $\chi_{yxz}$ & $\chi_{yxz}^*$ & $\chi_{xyz}^*$  \\
			\colrule
			$\chi_{xxz}$ & $-\chi_{yyz}$ & $\chi_{xxz}$ & $-\chi_{xxz}^*$ & $\chi_{yyz}^*$  \\
						\colrule
			$\chi_{yyz}$ & $-\chi_{xxz}$ & $\chi_{yyz}$ & $-\chi_{yyz}^*$ & $\chi_{xxz}^*$  \\
		\end{tabular}
	\end{ruledtabular}
	\label{susc2}
\end{table} 

\begin{table}[ht!]
	\caption{Transformation properties of the relevant susceptibilities for $\overline{4}'m'2$.}
	\centering
	\begin{ruledtabular}
		\begin{tabular}{c|c|c|c|c}
			$\overline{4}'m'2$ & $2TS_4$ & $C_{2z}$ & $2C_{2x}$ & $2T\sigma_d$  \\
			\colrule
			$\chi_{xyz}$ & $\chi_{yxz}^*$ & $\chi_{xyz}$ & $\chi_{xyz}$ & $\chi_{yxz}^*$  \\
						\colrule
			$\chi_{yxz}$ & $\chi_{xyz}^*$ & $\chi_{yxz}$ & $\chi_{yxz}$ & $\chi_{xyz}^*$  \\
			\colrule
			$\chi_{xxz}$ & $-\chi_{yyz}^*$ & $\chi_{xxz}$ & $-\chi_{xxz}$ & $\chi_{yyz}^*$  \\
						\colrule
			$\chi_{yyz}$ & $-\chi_{xxz}^*$ & $\chi_{yyz}$ & $-\chi_{yyz}$ & $\chi_{xxz}^*$  \\
		\end{tabular}
	\end{ruledtabular}
	\label{susc3}
\end{table} 

The previous relations are valid only for polar cartesian tensors (like the SHG susceptibility $\chi_{ijk}$), as they are based on the relations $\hat{S}_{4z}(x,y,z)\rightarrow(y,-x,-z)$ and $\hat{\sigma}_d(x,y,z)\rightarrow(y,x,z)$.
Their interpretation goes as follows: for $\overline{4}m'2'$, Table \ref{susc2} shows that there are two independent totally symmetric irreps, a non-magnetic and a magnetic one: $\Re (\chi_{xyz}+\chi_{yxz})$ and $\Im (\chi_{xxz}-\chi_{yyz})$. This also implies that $\Re (\chi_{xyz}-\chi_{yxz})=0$, and $\Im \chi_{xyz}=\Im \chi_{yxz}=0$ and that $\Im (\chi_{xxz}+\chi_{yyz})=0$, and $\Re \chi_{xxz}=\Re \chi_{yyz}=0$.
The non-magnetic $A_{1g}$ irrep, $\Re (\chi_{xyz}+\chi_{yxz})$, cannot be associated with the quadrupole polarization $\tilde{O}_{3z^2-r^2}$ of the first line of Eq.~(\ref{magquadcart}), as the latter is proportional to $\Re (\chi_{xyz}-\chi_{yxz})$ ($=0$). It can lead however to a signal from the octupole term $\tilde{O}_{xyz}$ of Eq.~(\ref{A2uoct}), proportional to $\Re (\chi_{xyz}+\chi_{yxz})$. Yet, the latter would give a signal in the SS channel, from Eq.~(\ref{azscanphi}), in contradiction with experiment \cite{harter}. So, the non-magnetic totally symmetric irrep of $\overline{4}m'2'$ can be excluded, as expected, because it coincides with the $\overline{4}m2$ case.
The magnetic irrep $\Im (\chi_{xxz}-\chi_{yyz})$, a magnetic quadrupole, is associated with the correct $T_{2u}^-$ symmetry of the primary OP that couples to the $\tilde{O}_{xy}$ polarization term in Eq.~(\ref{magquadcart}). However, the same irrep also couples to the magnetic octupole term $\tilde{O}_{z(x^2-y^2)}$ in Eq.~(\ref{T2uoct}), with the wrong angular dependence in the SP case and a non-zero SS signal. In order to remove this term, the extra condition $\Im (2\chi_{xxz}+\chi_{zxx})=0$ must be fulfilled, analogous to the proposal of Harter {\it et al.}~\cite{harter}.

For the $\overline{4}'m'2$ magnetic group, from Table \ref{susc3}, we get as well one magnetic and one non-magnetic totally symmetric irrep. In this case, $\chi_{xxz}=\chi_{yyz}=0$. The totally symmetric irrep is the linear combination $\chi_{xyz}+\chi_{yxz}^*$, with a magnetic and a non-magnetic contribution given by $\Im (\chi_{xyz}-\chi_{yxz})$ and $\Re (\chi_{xyz}+\chi_{yxz})$, respectively. This also implies that $\Re (\chi_{xyz}-\chi_{yxz})=0$, and $\Im (\chi_{xyz}+\chi_{yxz})=0$. We consider only the magnetic totally symmetric irrep $\Im (\chi_{xyz}-\chi_{yxz})$, as the non-magnetic case would again lead to the same conclusion of the $\overline{4}m2$ case. The term $\Im (\chi_{xyz}-\chi_{yxz})$ only couples to the polarization term $\tilde{O}_{3z^2-r^2}$ of Eq.~(\ref{magquadcart}). Remarkably, the coupling of this magnetic OP with the octupole term, proportional to  $\Im (\chi_{xyz}+\chi_{yxz})$, is automatically zero for $\overline{4}'m'2$ magnetic point group. There is no need to impose extra conditions like the ones imposed for the previous $\overline{4}m'2'$ group ($\Im (2\chi_{xxz}+\chi_{zxx})=0$) and for the $\overline{4}$ group ($2\chi_{xxz}-\chi_{zxx}=0$) \cite{harter}. So, all the symmetry conditions needed to obtain the experimental azimuthal scan are automatically satisfied in the $\overline{4}'m'2$ magnetic group, what makes it the most serious candidate to explain all the details of the SHG experiment.
This is confirmed by explicit calculations with a specific magnetic pattern satisfying the $\overline{4}'m'2$ symmetry, proposed in Section III.B.

We remark that all the previous terms are compatible with the measured temperature dependence (linear in $T_{c1}-T$ for secondary OPs and $\sqrt{T_{c1}-T}$ for primary OPs) and the constraints imposed by Landau theory \cite{harter}.
In the case of the $\overline{4}m'2'$ magnetic point group, the OPs involved are $E_u^+$, the secondary OP of $x^2-y^2$ symmetry, time-reversal even (the axial toroidal quadrupole) described in the previous subsection, $T_{2u}^-$, the primary OP that couples to the $\tilde{O}_{xy}$ polarization term, a magnetic quadrupole, and $T_{1g}^-$, a magnetic octupole of $z(5z^2-3r^2)$ symmetry (not contributing to the SHG signal, being inversion-even). All these OPs have the correct symmetry as required by the analysis already performed by Harter {\it et al.}~\cite{harter}, and therefore provide the correct temperature dependences for the secondary and primary OPs. A detailed analysis of this case, with a possible microscopic mechanism for its realization, is presented in the next subsection.
In the case of the $\overline{4}'m'2$ magnetic point group, the symmetries of the OPs involved are different: the primary/secondary OP coupling is of the kind $E_u^- E_u^+ A_{2g}^-$ that is an allowed term in the free energy, as the totally symmetric irrep $A_{1g}$ is contained in the product. Here, $E_u^-$ is a primary magnetic quadrupole of symmetry $M_{3z^2-r^2}$, time-reversal odd ($^-$), and parity-odd ($_u$), $E_u^+$ is the same secondary OP as above, an axial toroidal quadrupole of $x^2-y^2$ symmetry, time-reversal even ($^+$), and $A_{2g}^-$ is an $xyz$ component of the magnetic octupole, a primary OP sharing the same physical origin as the magnetic quadrupole $E_u^-$. A possible physical realization of this state is described in the next subsection. We conclude by highlighting the following technical remark. Rule 2 of Harter {\it et al.}'s SM8 \cite{harter} seems to exclude the possibility of having a primary OP transforming like a doublet $E_u$, as it forbids a linear coupling of the primary and secondary OPs in the free energy that would lead to a $\sqrt{T_{c1}-T}$ dependence for the secondary OP as well. However, this rule only applies to non-magnetic OPs: this coupling is automatically forbidden in our case, because it is of the kind $E_u^- E_u^+$, and would not be allowed in the free energy since it is time-reversal odd.

\subsection{Possible magnetic patterns for Cd$_2$Re$_2$O$_7$}

Here, we provide a possible physical realization of the two magnetic groups, $\overline{4}m'2'$ and $\overline{4}'m'2$.

\begin{figure}[ht!]
\includegraphics[width=0.45\textwidth]{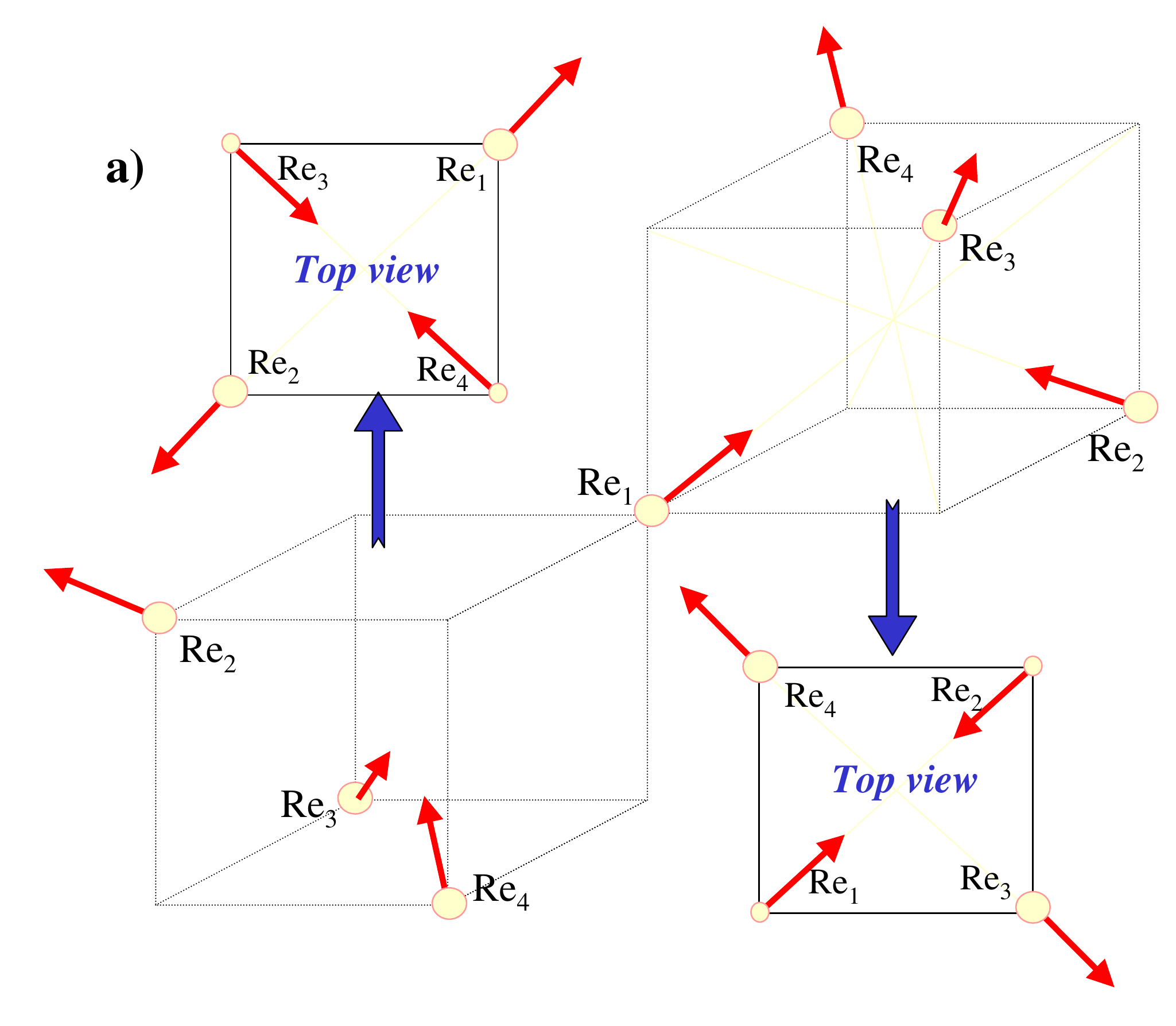}
\includegraphics[width=0.45\textwidth]{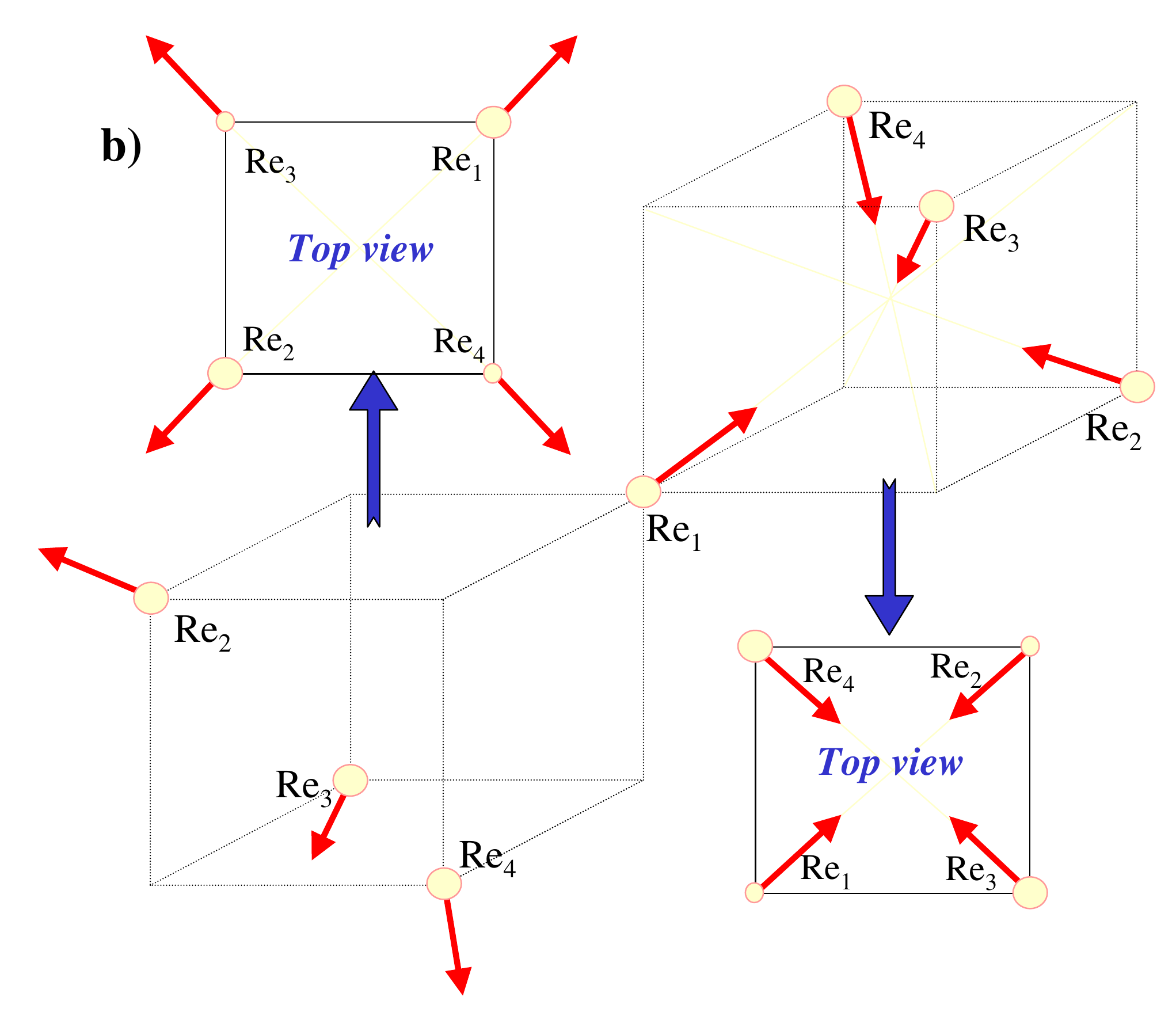}
\caption {(a) Possible magnetic pattern for $\overline{4}m'2'$ magnetic group, with its projection in the $xy$-plane, showing a two-in, two-out pattern. (b) Possible magnetic pattern for $\overline{4}'m'2$ magnetic group, with its projection in the $xy$-plane, showing an all-in all-out pattern.}
\label{magpat}
\end{figure}

\begin{figure*}[ht!]
\includegraphics[width=0.8\textwidth]{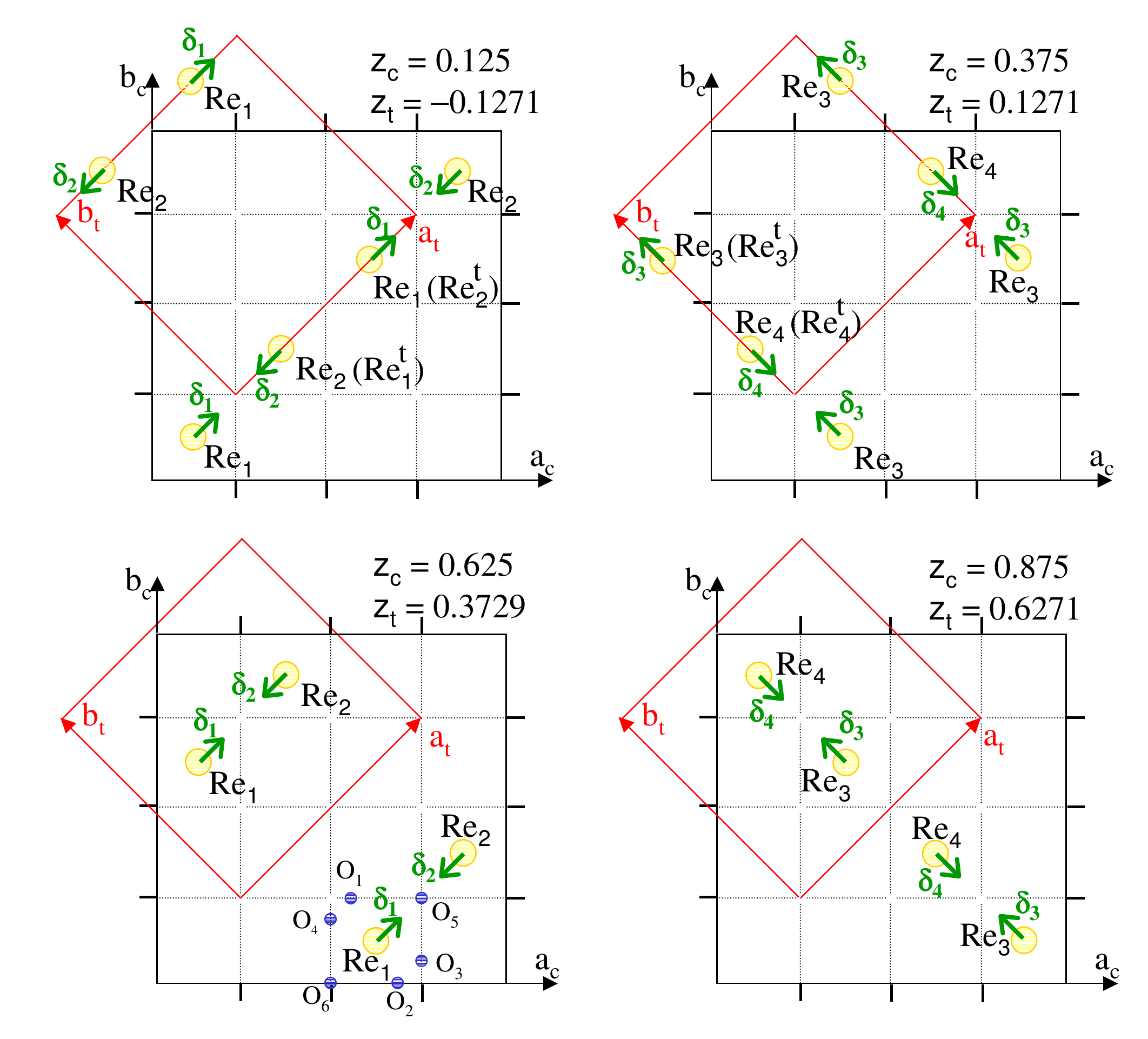}
\caption {Pyrochlore structure of Re ions depicted for the four planes along the $c$-axis. Green arrows represent in-plane Re displacements $\vec{\delta}_i$. Tetragonal $xy$-unit cell is shown in red. The $z$ coordinate of each plane is reported for the cubic cell ($z_c$) and for the tetragonal cell ($z_t$). Re$_1$ of the cubic cell (ITC No.~227) corresponds to Re$_2$ of the tetragonal cell (ITC No.~119). The 6 O ions around Re$_1$ are reported in blue. Their positions relative to Re$_1$ are given in Appendix C. We notice that O$_1$ and O$_3$ have both $z_c=3/4$.}
\label{pyrochlore}
\end{figure*}

{\it Case of $\overline{4}m'2'$:} The $\overline{4}m'2'$ group corresponds to the following magnetic configuration on a tetrahedron of Re ions (see Fig.~\ref{magpat}a):

\begin{align} \label{magdip2}
& \vec{m}_1=(\frac{\sin\alpha}{\sqrt{2}},\frac{\sin\alpha}{\sqrt{2}},\cos\alpha) \\
& \vec{m}_2=(-\frac{\sin\alpha}{\sqrt{2}},-\frac{\sin\alpha}{\sqrt{2}},\cos\alpha) \nonumber \\
& \vec{m}_3=(\frac{\sin\alpha}{\sqrt{2}},-\frac{\sin\alpha}{\sqrt{2}},\cos\alpha) \nonumber \\
& \vec{m}_4=(-\frac{\sin\alpha}{\sqrt{2}},\frac{\sin\alpha}{\sqrt{2}},\cos\alpha) \nonumber 
\end{align}
corresponding to the Re ions of Eq.~(\ref{positions}). Here $\alpha$ is the angle between the magnetic moment and the $c$-axis. The origin is taken at the center of the tetrahedron, inscribed in a cube of unit length.
We remind that the group $\overline{4}m2$ is non-abelian as $\hat{S}_{4z}\hat{C}_{2x}\neq \hat{C}_{2x}\hat{S}_{4z}$. This implies that the labeling of the Re atoms on the tetrahedron must take into account the correct space-group symmetry relations of the tetragonal $I\overline{4}m2$ crystal, as the Re$_i$ are not translationally equivalent, not even in the cubic phase (see Fig.~\ref{pyrochlore}).
The total magnetic moment in a tetragonal unit cell (made of the two identical tetrahedra) is zero only if $\alpha=\pi/2$. In this case, the magnetic quadrupole is independent of the choice of the origin. Keeping the center of the tetrahedron as the origin, we get the five components of the magnetic quadrupole per Re ion:
\begin{align}
& M_{xy}\equiv \frac{1}{4} \sum_{i=1}^4 r_{ix}m_{iy}+r_{iy}m_{ix} = \frac{\sin\alpha}{4\sqrt{2}} \nonumber \\
& M_{xz}\equiv \frac{1}{4} \sum_{i=1}^4 r_{ix}m_{iz}+r_{iz}m_{ix} = 0 \nonumber \\
& M_{yz}\equiv \frac{1}{4} \sum_{i=1}^4 r_{iz}m_{iy}+r_{iy}m_{iz} = 0 \nonumber \\
& M_{x^2-y^2}\equiv \frac{1}{4} \sum_{i=1}^4 r_{ix}m_{ix}-r_{iy}m_{iy} = 0 \nonumber \\
& M_{3z^2-r^2}\equiv \frac{1}{4} \sum_{i=1}^4 3r_{iz}m_{iz}-\vec{r}\cdot\vec{m} = 0 
\end{align}
Here $r_{ix}$ is the $x$-component of the Re$_i$ atom, etc. The only non-zero  component, of $T_{2u}$ symmetry, combines with $\tilde{O}_{xy}$ polarization.
For future use, we remark that the equivalent components of the toroidal magnetic quadrupole (defined with $\vec{t}_i=\vec{r}_i\times\vec{m}_i$ replacing $\vec{m}_i$ in all previous formulas) are zero. The three relevant components of the magnetic octupole are (the remaining four are zero):
\begin{align}
& O_{xyz}\equiv \frac{1}{4} \sum_{i=1}^4 r_{ix}r_{iy}m_{iz}+r_{ix}m_{iy}r_{iz}+ m_{ix}r_{iy}r_{iz}= 0 \nonumber \\
& O_{z(x^2-y^2)}\equiv \frac{1}{4} \sum_{i=1}^4 m_{iz}(r_{ix}^2-r_{iy}^2) \nonumber \\
& \hspace{0.75in} +2r_{iz}(r_{ix}m_{ix}-r_{iy}m_{iy}) = 0 \nonumber \\
& O_{z(5z^2-3r^2)}\equiv \frac{1}{4} \sum_{i=1}^4 m_{iz}(5r_{iz}^2-r_{i}^2) \nonumber \\
& \hspace{0.25in} +2r_{iz}(5r_{iz}m_{iz}-3\vec{r}_i\cdot\vec{m}_i) = \frac{3}{32}(\cos\alpha-\sqrt{2}\sin\alpha) \nonumber
\end{align}
where the first term above has $A_{2g}$ symmetry, the second $T_{2g}$ symmetry, and the last $T_{1g}$ symmetry relative to the (001) direction. This last term has the same physical origin as the magnetic quadrupole term, $M_{xy}$, so that the two can represent the coupled $T_{1g}$-$T_{2u}$ terms needed for the Landau symmetry analysis given in Harter {\it et al.}~\cite{harter}. We remind that the $T_{1g}$ octupole component cannot be seen directly by SHG, at least as a purely electric-dipole (E1-E1-E1) signal, since it is inversion even. 

There is, however, a drawback associated with the $I\overline{4}m'2'$ magnetic point group: if we evaluate the toroidal octupole (inversion odd) associated with this magnetic configuration, we obtain the following nonzero component (the remaining six, in particular $T_{xyz}$, are zero):
\begin{align}
&T_{z(x^2-y^2)}\equiv \frac{1}{4} \sum_{i=1}^4 t_{iz}(r_{ix}^2-r_{iy}^2)+2r_{iz}(r_{ix}t_{ix}-r_{iy}t_{iy}) \nonumber \\
&\hspace{0.75in} = \frac{1}{128}(\cos\alpha-\frac{\sin\alpha}{\sqrt{2}}) \nonumber
\end{align}

This nonzero component has $T_{2u}$ symmetry and couples to the octupolar polarization terms $\tilde{O}_{z(x^2-y^2)}^{SP}$ and $\tilde{O}_{z(x^2-y^2)}^{SS}$ evaluated in Eq.~(\ref{azscanphi}) for the SHG signal. In order to reproduce the correct azimuthal SP scan and the absence of an SS signal, they should have been instead zero (this corresponds to the relation $\chi_{zxx}=-2\chi_{xxz}$ imposed by Harter {\it et al.}). 
$T_{z(x^2-y^2)}$ is zero only for $\tan\alpha=\sqrt{2}$, i.e., $\alpha\simeq 54.74^\circ$, corresponding to the magnetic moment along the trigonal axis. This result, as for all the others in this subsection, has been calculated for the cubic positions $\vec{r}_i$. However, $T_{z(x^2-y^2)}$ remains zero for this value of $\alpha$ even with a tetragonal distortion. In fact, even allowing $\vec{r}_i\rightarrow \vec{r}_i+\vec{\delta}_i$, the change in $T_{z(x^2-y^2)}$ is $\Delta T_{z(x^2-y^2)} = (\frac{x_1+z_1}{16}+\frac{x_1z_1}{2})(\cos\alpha-\frac{\sin\alpha}{\sqrt{2}})$.
We remark that the other relation imposed by Harter {\it et al.}, $\chi_{zxy}=-2\chi_{xyz}$, is automatically satisfied, as it corresponds to the $A_{2u}$ term, $T_{xyz}$, which is null. 

This confirms the general symmetry analysis of Section III.A with the important addition that for $\alpha\simeq 54.74^\circ$, even with a tetragonal distortion, this magnetic group could explain the Harter {\it et al.}'s condition $\chi_{zxx}=-2\chi_{xxz}$. Yet, just having a single angle where the condition is fulfilled seems implausible. Moreover, this solution would be ferromagnetic, against experiment (only $\alpha=\pi/2$ would lead to an acceptable antiferromagnetic solution). Interestingly, the above relations for $\chi_{zxx}$ and $\chi_{zxy}$ are instead both automatically satisfied for any angle $\alpha$ in the model presented below, corresponding to the $\overline{4}'m'2$ group. This leads to a null SS signal, independent of any other imposed constraint. 

{\it Case of $\overline{4}'m'2$:} 
Magnetic moments are compatible with the $I\overline{4}'m'2$ magnetic group if they have the configuration (see Fig.~\ref{magpat}b):

\begin{align} \label{magdip}
& \vec{m}_1=(\frac{\sin\alpha}{\sqrt{2}},\frac{\sin\alpha}{\sqrt{2}},\cos\alpha) \\
& \vec{m}_2=(-\frac{\sin\alpha}{\sqrt{2}},-\frac{\sin\alpha}{\sqrt{2}},\cos\alpha) \nonumber \\
& \vec{m}_3=(-\frac{\sin\alpha}{\sqrt{2}},\frac{\sin\alpha}{\sqrt{2}},-\cos\alpha) \nonumber \\
& \vec{m}_4=(\frac{\sin\alpha}{\sqrt{2}},-\frac{\sin\alpha}{\sqrt{2}},-\cos\alpha) \nonumber 
\end{align}

We remark that, as in Eq.~(\ref{magdip2}), these magnetic moments have to stay within the local mirror planes at sites Re$_i$, so as to enforce the $m'$ symmetry. Atoms Re$_1$ and Re$_4$ must be related by the two-fold axis around $z$ (as well as Re$_2$ and Re$_3$). Finally, $\vec{m}_2=\hat{T}\hat{S}_{4z}\vec{m}_1$ and $\vec{m}_3=\hat{T}\hat{S}_{4z}^3\vec{m}_1$. This configuration has $I\overline{4}'m'2$ symmetry for any $\alpha$, which is the angle that the magnetic dipole makes with the $c$-axis. The total magnetic dipole of the tetrahedron is zero.  

We can now evaluate the magnetic quadrupole and octupole components for both the cubic and tetragonal phases.
In the cubic phase, we get that all magnetic quadrupole components are identically zero, except $M_{3z^2-r^2}\propto \cos\alpha -\sin\alpha/\sqrt{2}$.
Analogously, for the octupole magnetic moments, six components are zero ($O_{z(x^2-y^2)}$, $O_{x(y^2-z^2)}$, $O_{y(z^2-x^2)}$, $O_{z(5z^2-3r^2)}$, $O_{x(5x^2-3r^2)}$, $O_{y(5y^2-3r^2)}$) and only one is different from zero, the $A_{2g}$ term, that is given by: $O_{xyz}\propto \cos\alpha +\sqrt{2} \sin\alpha$. 
We remark that the magnetic quadrupole $M_{3z^2-r^2}$ is zero when the magnetic octupole $O_{xyz}$ is maximal, i.e., for $\alpha=54.74^\circ$, corresponding to magnetic dipoles along the three-fold axis of the cube diagonal. Therefore, in the high-temperature cubic phase where the three-fold symmetry is present, the magnetic configuration of Eq.~(\ref{magdip}) is equivalent to the all-in all-out magnetic pattern of Cd$_2$Os$_2$O$_7$ \cite{arima}, with zero magnetic quadrupole and the $O_{xyz}$ magnetic octupole as the lowest non-zero magnetic multipole.

We now evaluate the same quantities in the $I\overline{4}m2$ phase, where $\vec{r}_i \rightarrow \vec{r}_i+\vec{\delta}_i$, in order to identify a possible relation between the magnetic quadrupole and octupole and the tetragonal distortion. The displacements from the cubic positions for each Re atom in the two tetragonal phases, reported above in Eq.~(\ref{displs}), can be rewritten for the $I\overline{4}'m'2$ magnetic group ($x_1=y_1$) as: $\vec{\delta}_1=(x_1,x_1,z_1)$, $\vec{\delta}_2=(x_1,-x_1,-z_1)$, $\vec{\delta}_3=(-x_1,x_1,-z_1)$ and $\vec{\delta}_4=(-x_1,-x_1,z_1)$.
 Then we get that again $M_{xz}=M_{yz}=M_{xy}=M_{x^2-y^2}=0$ and $O_{z(x^2-y^2)}=O_{x(y^2-z^2)}=O_{y(z^2-x^2)}=O_{z(5z^2-3r^2)}=O_{x(5x^2-3r^2)}=O_{y(5y^2-3r^2)}=0$. The changes $\Delta M_{3z^2-r^2}$ and $\Delta O_{xyz}$ in the terms $M_{3z^2-r^2}$ and $O_{xyz}$ between the tetragonal and the cubic phases, up to the first order in the displacement terms, are given by:

\begin{align} 
\Delta M_{3z^2-r^2} = 2 z_1 \cos\alpha -\sqrt{2} x_1 \sin\alpha \nonumber \\
\Delta O_{xyz} = \frac{x_1}{4} \cos\alpha +\frac{\sin\alpha}{4\sqrt{2}} (x_1+z_1)
\label{magoctquad}
\end{align}

We remark that the tetragonal correction to the magnetic quadrupole due to the displacements $\vec{\delta}_i$ is not zero even when the magnetic moment is directed along the local three-fold axes.
In this sense, Eq.~(\ref{magoctquad}) highlights the connection between a non-zero value of the magnetic quadrupole and the tetragonal distortion. For $x_1$ and $z_1$ around the experimental values ($x_1\sim 0.002$ and $z_1\sim -0.002$, in fractional units), the correction determined by Eq.~(\ref{magoctquad}) is almost constant for small variations of $\alpha$ (as $\Delta M_{3z^2-r^2} \propto \sqrt{2} \cos\alpha + \sin\alpha$ is a maximum for $\alpha=54.74^\circ$).

Interestingly, all toroidal octupole components associated with this magnetic configuration are zero and, in particular, $T_{z(x^2-y^2)}=0$ and $T_{xyz}=0$, i.e., the only two components that would have both given a spurious signal in the SS channel and the wrong azimuthal scan in the SP channel, as detailed in Section II.B. So this solution is fully compatible with Harter {\it et al.}'s azimuthal scan, as the constraints $\chi_{zxx}=-2\chi_{xxz}$ and $\chi_{zxy}=-2\chi_{xyz}$ are automatically satisfied by this magnetic pattern, in keeping with the general symmetry analysis of Section III.A. This is a remarkable result that suggests that $\overline{4}'m'2$ is the actual magnetic point group.
Moreover, as sketched above, the non-zero component $M_{3z^2-r^2}$ has $E_u^-$ symmetry. The only non-zero component of the magnetic octupole generated by this configuration, $O_{xyz}$, has $A_{2g}$ symmetry and can represent the even-parity primary OP in the Landau free energy. So, all OPs are correctly represented by the non-zero magnetic multipoles associated with this case.

In the next section, we suggest new REXS experiments that could highlight the magnetic pattern of the $I\overline{4}'m'2$ magnetic space group.

\section{REXS analysis of the phase transitions in Cd$_2$Re$_2$O$_7$}

In this section, we analyze the possible outcomes of the predicted magnetic-quadrupole ground state by means of resonant X-ray elastic scattering (REXS), that has the sensitivity to confirm the proposed magnetic configuration. We shall also touch on the possible signature of the other phase transition reported in the literature, around T=120 K, again investigated by REXS.

\subsection{Probing by REXS the $T_{c1}$ phase transition of ${\rm Cd_2Re_2O_7}$: how to confirm a magnetic-quadrupole OP}

In Table \ref{sym1} we report the fractional coordinates of Re atoms for both $Fd\overline{3}m$ and $I\overline{4}m2$ space groups as measured by Huang {\it et al.}~\cite{huang}. In keeping with the previous sections, we have switched the origin choice 2 used by Huang {\it et al.}~for $Fd\overline{3}m$ to the origin choice 1, corresponding to a global shift of (1/8,1/8,1/8) for all the atoms. We remark also that with our choice of Re$_1$ for the cubic phase, the corresponding Re atom in the tetragonal phase is the second entry of the ITC, No.~119. This point can be checked from the symmetry operations reported in the ITC (Nos.~119 and 227), by reminding that the group is not abelian (as the ${\hat{S}}_{4z}$ operator does not commute with two-fold rotations, the order is important): ${\hat{S}}_{4z}{\hat{C}}_{2x\bar{x}}={\hat{C}}_{2x{x}}{\hat{S}}_{4z}={\hat{\sigma}}_{x}$, whereas ${\hat{S}}_{4z}{\hat{C}}_{2x{x}}={\hat{C}}_{2x\bar{x}}{\hat{S}}_{4z}={\hat{\sigma}}_{y}$. So, the correct correspondence is Re$_1\rightarrow {\rm Re}_2^t$; Re$_2\rightarrow {\rm Re}_1^t$; Re$_3\rightarrow {\rm Re}_3^t$; Re$_4\rightarrow {\rm Re}_4^t$, as shown in Fig.~\ref{pyrochlore}.

\begin{table}[ht!]
	\caption{Fractional positions $(x,y,z)$ of Re atoms in the $Fd\overline{3}m$ and $I\overline{4}m2$ space groups. The twelve fcc translations (for $Fd\overline{3}m$) and the four bcc translations (for $I\overline{4}m2$) are not reported.  Of all symmetry operations relative to Re$_1$, only the one used to derive the resonant structure factor is highlighted.}
	\centering
	\begin{ruledtabular}
		\begin{tabular}{c|c|c|c|c}
			Atom & $(x,y,z)$~$Fd\overline{3}m$ & $Fd\overline{3}m$ & $(x,y,z)$~$I\overline{4}m2$ & $I\overline{4}m2$ \\
			\colrule
			Re$_1$ & ($\frac{1}{8}$,$\frac{1}{8}$,$\frac{1}{8}$) & ${\hat{E}}$ & (0.7529,0,0.8729) & ${\hat{C}}_{2z}$   \\
			Re$_2$ & ($-\frac{1}{8}$,$-\frac{1}{8}$,$\frac{1}{8}$) & ${\hat{C}}_{2z}$ & (0.2471,0,0.8729) &  ${\hat{E}}$ \\
			Re$_3$ & ($-\frac{1}{8}$,$\frac{1}{8}$,$-\frac{1}{8}$) & ${\hat{C}}_{2y}$ & (0,0.7529,0.1271) & ${\hat{I}}{\hat{C}}_{4z}$  \\
			Re$_4$ & ($\frac{1}{8}$,$-\frac{1}{8}$,$-\frac{1}{8}$) & ${\hat{C}}_{2x}$ & (0,0.2471,0.1271) & ${\hat{I}}{\hat{C}}_{4z}^-$ \\
		\end{tabular}
	\end{ruledtabular}
	\label{sym1}
\end{table} 

Here and in what follows, ${\hat{I}}$ and ${\hat{E}}$ are the inversion and the identity operators, and ${\hat{C}}_{2i}$ are two-fold rotations around the $i$ axis ($x,y,z$ parallel to cubic $a,b,c$). In the following, we label $w=0.2471$ and $u=0.8729$ \cite{huang}. Taking into account the space-group symmetries, the resonant X-ray structure factor for the $Fd\overline{3}m$ space group, summed over the 16 Re atoms, can be written as: 

\begin{align} \label{strfac3}
& F^{hkl}_{Fd\overline{3}m} \propto  (1+(-1)^{h+k}+(-1)^{h+l}+(-1)^{k+l}) \\
& \hspace{0.5in} (1+(i)^{h+k}{\hat{C}}_{2z} +(i)^{h+l}{\hat{C}}_{2y} +(i)^{k+l}{\hat{C}}_{2x})f_1 \nonumber
\end{align}
where $f_1$ is the resonant atomic scattering amplitude for Re atom 1 (see, e.g., Ref.~\cite{jphysD}). 

The resonant X-ray structure factor for the  $I\overline{4}m2$ space group, summed over the 8 Re atoms, can be written as: 

\begin{align}
& F^{hkl}_{I\overline{4}m2} \propto (1+(-1)^{h+k+l})(e^{2i\pi (hw+lu)}(1+e^{-4i\pi hw}{\hat{C}}_{2z}) \nonumber \\
& \hspace{0.5in} +e^{-2i\pi (kw+lu)}(1+e^{+4i\pi kw}{\hat{C}}_{2z}){\hat{I}}{\hat{C}}_{4z})f_{1t}
\label{strfac1}
\end{align}
where $f_{1t}$ is the resonant atomic scattering amplitude for tetragonal $Re_1^t$.
We remark that for the all-in, all-out $I\overline{4}'m'2$ magnetic group, the only change in the structure factor is the replacement of ${\hat{I}}{\hat{C}}_{4z}$ in the last term by ${\hat{T}}{\hat{I}}{\hat{C}}_{4z}$. 

\begin{figure}[ht!]
\includegraphics[width=0.45\textwidth]{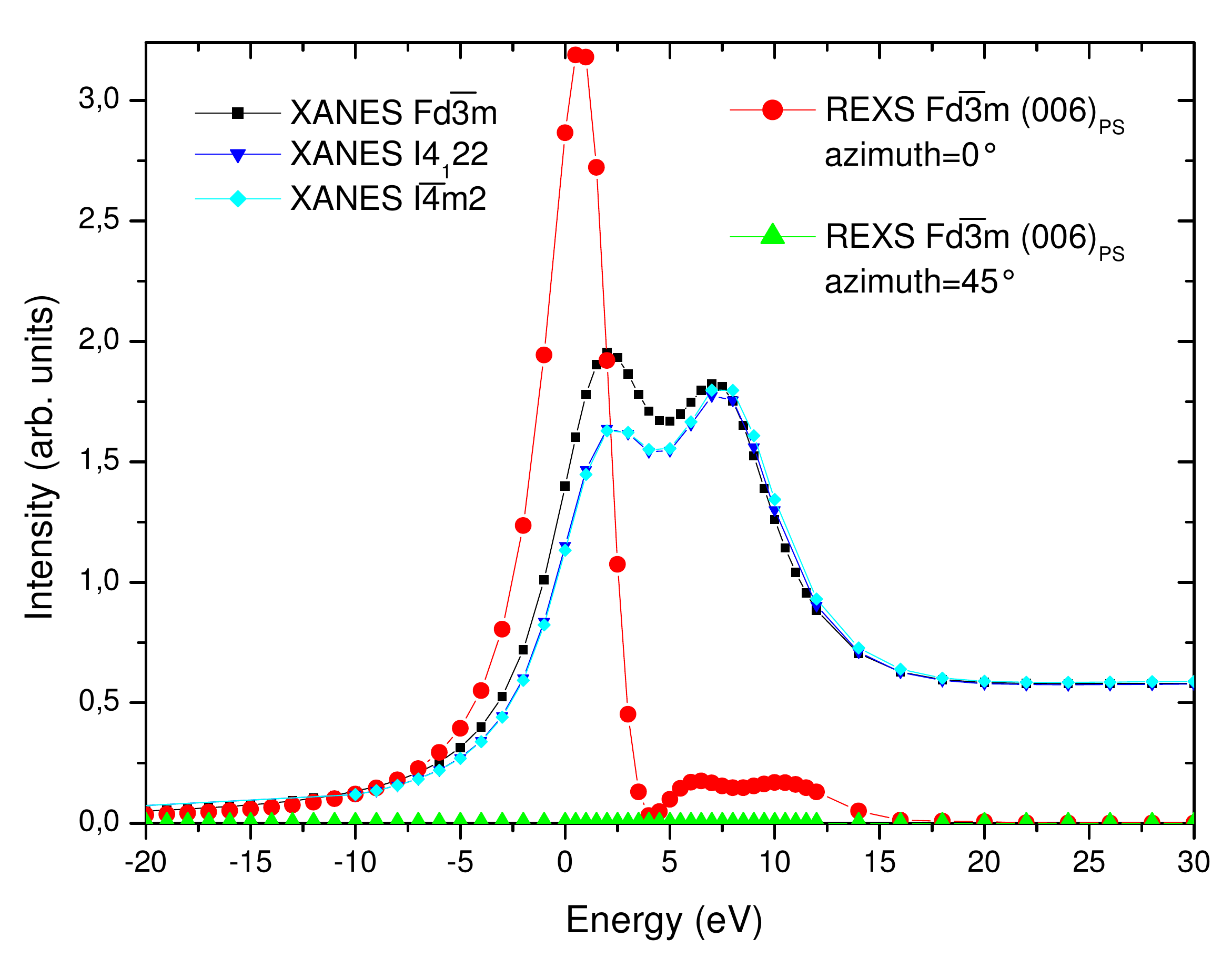}
\caption {The X-ray absorption intensity (XAS) near the Re L$_3$ edge for the $Fd\overline{3}m$ and ${I\overline{4}m2}$ space groups. The two peaks are due to the unoccupied t$_{2g}$ and e$_g$ states.  We also show the REXS signal for the (006) reflection in SP geometry, at both $0^\circ$ and $45^\circ$ azimuth. The latter signal is zero, suggesting where to look for the magnetic signal in the low-temperature phase.}
\label{xas}
\end{figure}

A key class of reflections that can be used in REXS to identify the breaking of the $Fd\overline{3}m$ space group is $(0,0,4n+2)$.
If we specialize Eq.~(\ref{strfac3}) in this case, we get:

\begin{align}
& F^{0,0,4n+2}_{Fd\overline{3}m} \propto 4 (1+{\hat{C}}_{2z} -{\hat{C}}_{2y} -{\hat{C}}_{2x})f_1
\label{strfac3bis}
\end{align}
which is forbidden out of resonance (it gives zero if we put ${\hat{C}}_{2z}={\hat{C}}_{2x}={\hat{C}}_{2y}=1$, as appropriate out of resonance). 
In anomalous conditions, though, such a reflection becomes allowed. 
We remind that, at L$_{2,3}$ edges, only E1-E1 transitions are possible and REXS is sensitive only to a time-reversal even OP, which is the electric quadrupole $Q$, and a time-reversal odd OP, the magnetic dipole $\vec{m}$. In the high-temperature cubic phase, there is no magnetism and only $Q$ can be detected at the (0,0,4n+2) reflections. Of the 5 components of $Q$ ($Q_{3z^2-r^2}$, $Q_{x^2-y^2}$, $Q_{xz}$, $Q_{yz}$, $Q_{xy}$), the only term that is nonzero when acted upon by the linear combination $(1+{\hat{C}}_{2z} -{\hat{C}}_{2y} -{\hat{C}}_{2x})$ of Eq.~(\ref{strfac3bis}) is $Q_{xy}$ (as ${\hat{C}}_{2z}Q_{xy}=+Q_{xy}$, ${\hat{C}}_{2y}Q_{xy}=-Q_{xy}$ and ${\hat{C}}_{2x}Q_{xy}=-Q_{xy}$). All the other terms are zero.

If we perform an azimuthal scan around $\vec{q}=(0,0,4n+2)$, and put the azimuth zero corresponding to the incoming S polarization (perpendicular to the scattering plane) along the $\vec{a}$-axis, then we get an azimuthal scan $\sim \cos^2(2\phi)$ for SP and PS scattering and $\sim \sin^2(2\phi)$ for SS and PP scattering, because of the $xy$ dependence for the only allowed term $Q_{xy}$. We remark that each P polarization further introduces a reduction by $\sin\theta$, where $\theta$ is the Bragg angle.
This azimuthal behavior for non-magnetic REXS is the same as obtained by Yamaura {\it et al.}~\cite{yamaura}, who studied the (006) reflection for Cd$_2$Os$_2$O$_7$. We finally remark that the $Q_{xy}$ electric quadrupole does not probe only Re $d_{xy}$ orbitals, because the local oxygen octahedron around each Re-ion is not oriented along the $\vec{a}$, $\vec{b}$, and $\vec{c}$ cubic axes. The calculation to rotate the local axes to the crystal axes is performed in Appendix C. The result shows that in the energy scan of the (0,0,4n+2) reflections, both $t_{2g}$ and $e_g$ local states are detected with a relative weight of $65\%$ for $t_{2g}$ and $35\%$ for $e_{g}$. The $e_g$ peak is further smeared by lifetime broadening, which increases by a factor of about two when passing from the lower energy $t_{2g}$ states to the higher energy $e_g$ ones. This is confirmed by our FDMNES numerical calculations \cite{joly}, fully relativistic (spin-orbit included) in SP geometry for $\phi=0^\circ$ and $\phi=45^\circ$ shown in Fig.~\ref{xas} at the Re L$_3$ edge. We remark that similar results are obtained at the L$_2$ edge \cite{noteL23}. Huang {\it et al.}'s positions \cite{huang} for the $Fd\overline{3}m$ space group were used. As Cd$_2$Os$_2$O$_7$ is a metal, we did not include a Hubbard $U$, but we checked that nonzero values of $U$ (up to 2 eV for both Cd and Re) do not affect the main conclusions of this section.
As noted above, no signal appears for $\phi=45^\circ$. The $t_{2g}$-$e_g$ separation is about 5 eV, in keeping with the ab initio calculations of Huang {\it et al.}~\cite{huang}. 

Below the transition, in ${I\overline{4}'m'2}$ group, $(0,0,4n+2)$ reflections become instead Bragg-allowed. In fact, if we specialize Eq.~(\ref{strfac1}) to the case $(0,0,4n+2)$ for the magnetic group $I\overline{4}'m'2$, we have:

\begin{align}
F^{0,0,4n+2}_{I\overline{4}'m'2} \propto 2(1+{\hat{C}}_{2z})(e^{i\pi (8n+4)u}+e^{-i\pi (8n+4)u}{\hat{T}}{\hat{I}}{\hat{C}}_{4z})f_{1t}
\label{strfac1bis}
\end{align}

Though $l=4n+2$ reflections are Bragg-allowed, the Bragg term only contributes to SS or PP scattering, whereas the magnetic signal is only visible in SP or PS geometry, together with the electric-quadrupole terms, $Q$. However, in a real experiment, there is the further issue of possible leakage in the SP channel of Bragg scattering from the SS channel. 
Because of this, a full symmetry analysis for all the possible terms must be performed. Of the five electric-quadrupole components, $Q_{\tilde{x}z}$ and $Q_{\tilde{y}z}$ give exactly no signal, being odd with respect to ${\hat{C}}_{2z}$. The other three are even under ${\hat{C}}_{2z}$. $Q_{3z^2-r^2}$ is also even for ${\hat{T}}{\hat{I}}{\hat{C}}_{4z}$, whereas $Q_{\tilde{x}^2-\tilde{y}^2}$ and $Q_{\tilde{x}\tilde{y}}$ are odd, because of the ${\hat{C}}_{4z}$ rotation (all electric quadrupoles are even under ${\hat{T}}{\hat{I}}$). Therefore, $Q_{3z^2-r^2}$ is reduced by $\cos(2\pi u(4n+2))$ \cite{notevalues}, whereas $Q_{\tilde{x}^2-\tilde{y}^2}$ and $Q_{\tilde{x}\tilde{y}}$ go as $i \sin(2\pi u(4n+2))$. We remark that Eq.~(\ref{strfac1bis}) is written in terms of tetragonal symmetry operations, with $x$ and $y$ cubic and $\tilde{x}$ and $\tilde{y}$ tetragonal axes rotated by $45^\circ$ - see Fig.~\ref{pyrochlore}. So, $Q_{\tilde{x}^2-\tilde{y}^2}$ corresponds to $Q_{xy}$ of the cubic phase. A numerical calculation of the order of magnitude of these terms by the FDMNES program shows that $Q_{\tilde{x}\tilde{y}}$ and $Q_{3z^2-r^2}$, i.e., the components induced by the tetragonal distortion ($\propto \delta_i^2$) are totally negligible compared to $Q_{\tilde{x}^2-\tilde{y}^2}$, to the magnetic $m_z$ term, and to the Bragg term. If we fix to $1000$ the intensity of the Bragg term, $I_{Q_{3z^2-r^2}}\sim 10^{-2}$, $I_{Q_{\tilde{x}\tilde{y}}}\sim 10^{-4}$, $I_{Q_{\tilde{x}^2-\tilde{y}^2}}\sim 10$ and $I_{m_{z}}\sim 10$. As above, our calculations were fully relativistic, with spin-orbit included. Huang {\it et al.}'s atom positions \cite{huang} for the $I\overline{4}m2$ space group were used.
Therefore the total intensity can be simplified as follows ($l=4n+2$): 

\begin{align}
F^{00l}_{I\overline{4}'m'2}&=8\big[\cos(2l\pi u)(f_{Q_0}+f_{Q_{3z^2-r^2}})-\sin(2l\pi u)f_{m_z}\big]\nonumber \\
&+8i\sin(2l\pi u)\big[f_{Q_{\tilde{x}^2-\tilde{y}^2}}+f_{Q_{\tilde{x}\tilde{y}}}\big]  \nonumber \\
&\simeq 8\big[\cos(2l\pi u)f_{Q_0}-\sin(2l\pi u)f_{m_z}\big]  \nonumber \\
&+8i\sin(2l\pi u)f_{Q_{\tilde{x}^2-\tilde{y}^2}}
\label{strfacmag}
\end{align}
where we have explicitly written each multipole contribution to the atomic scattering factor $f_{1t}$, including the Bragg term, $f_{Q_0}$.

\begin{figure}[ht!]
\includegraphics[width=0.44\textwidth]{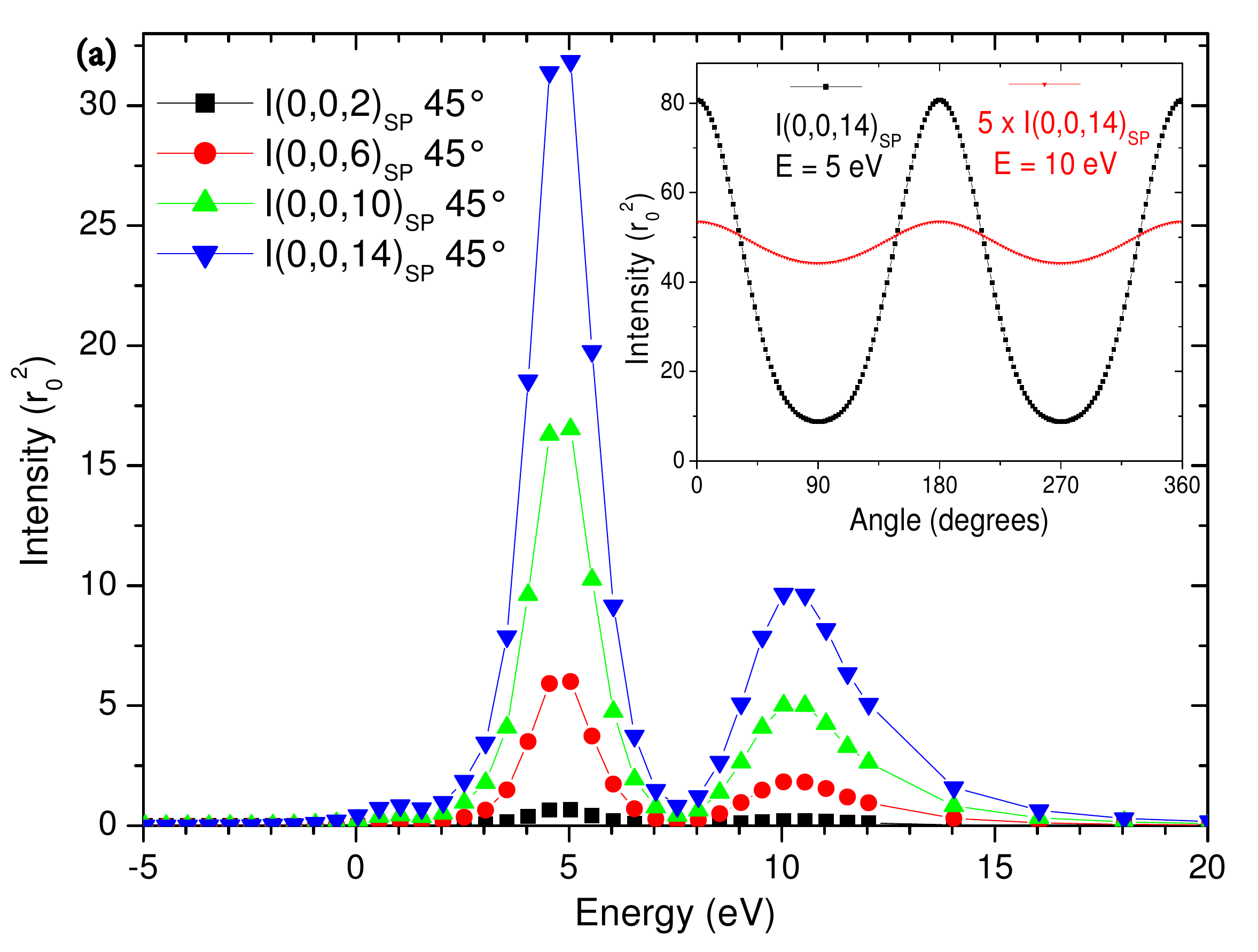}
\includegraphics[width=0.44\textwidth]{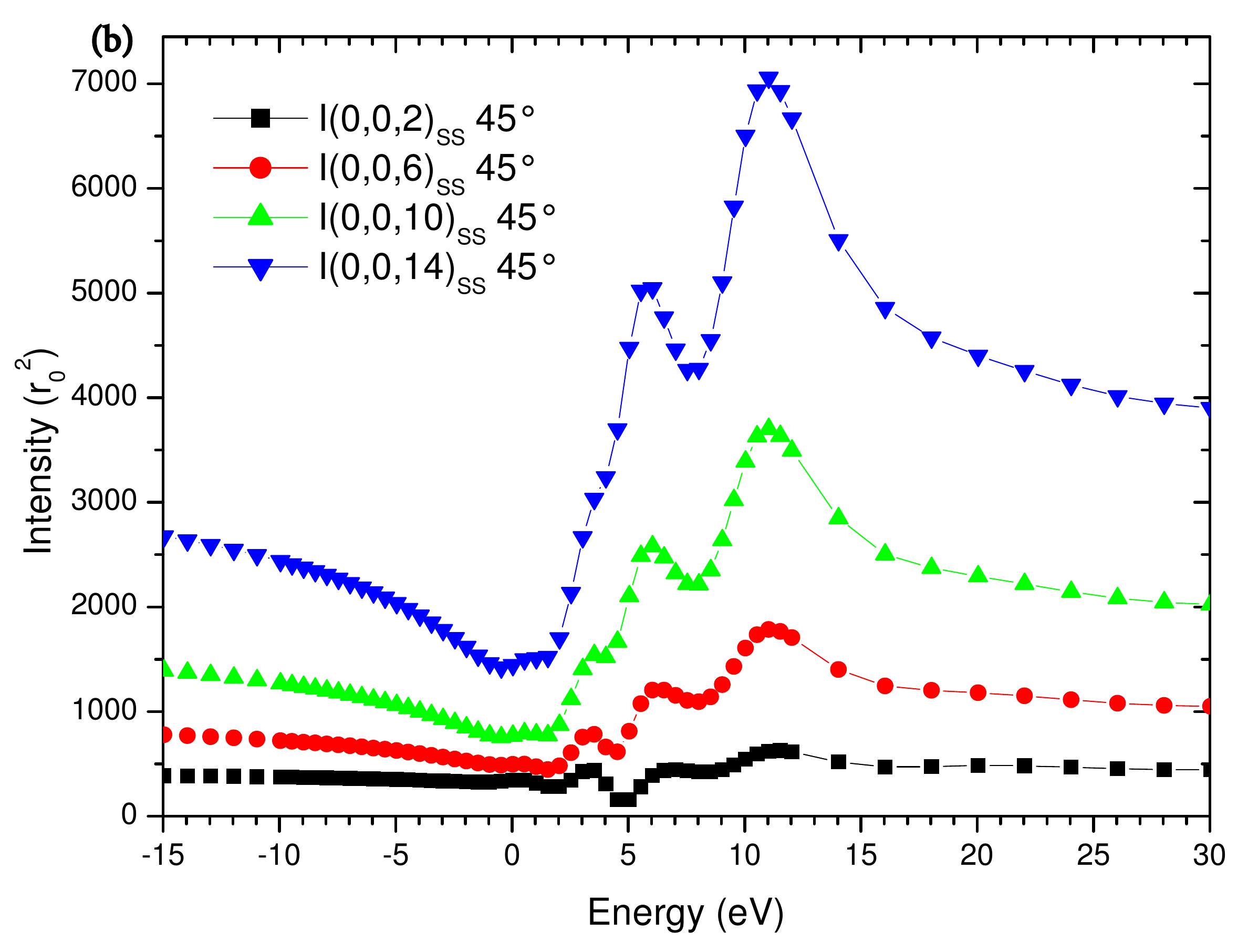}
\includegraphics[width=0.44\textwidth]{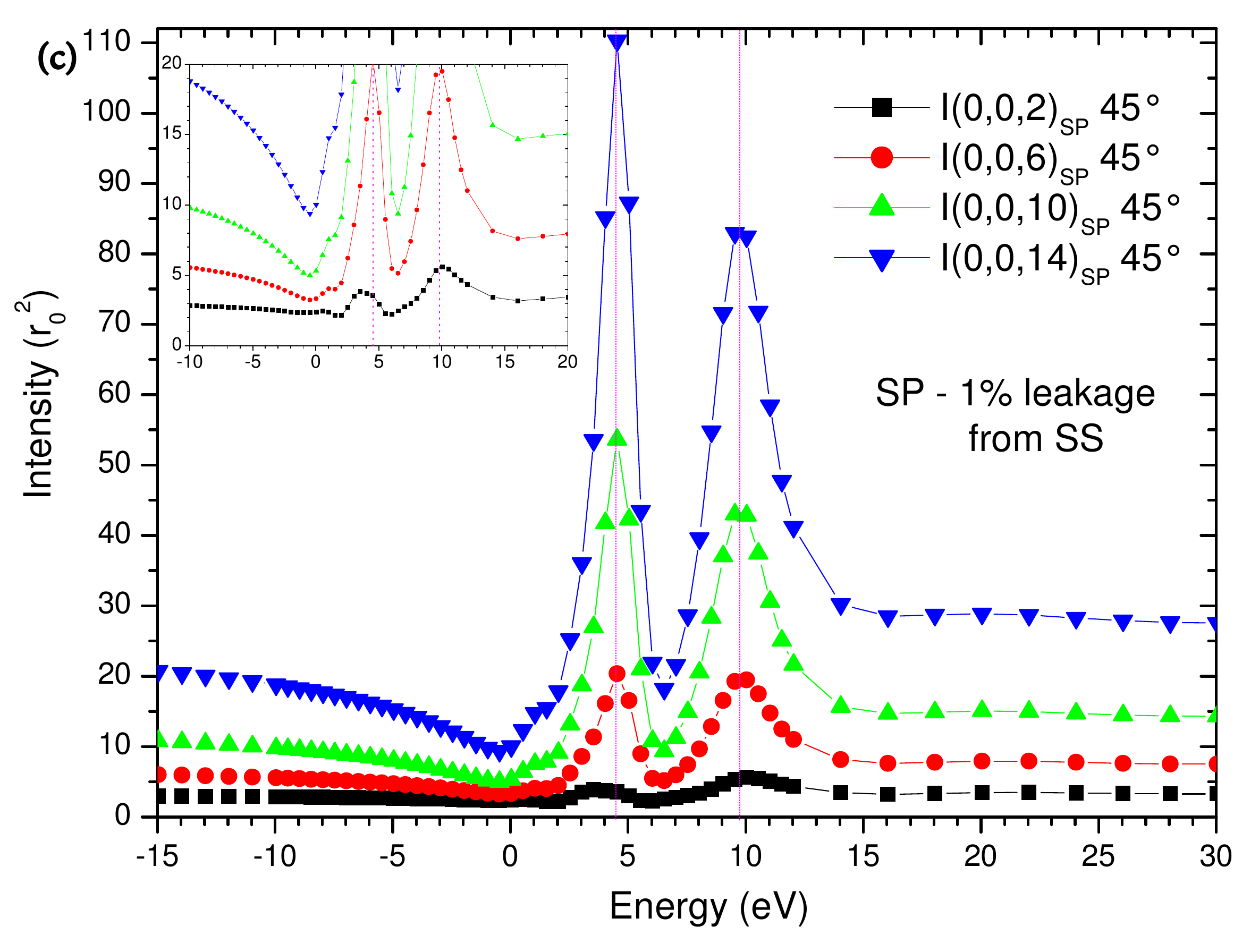}
\caption {a) Re L$_3$-edge SP $(0,0,4n+2)$ energy scans. The signal is purely magnetic, with a main contribution at the $t_{2g}$ energy and a smaller one at the $e_g$ energy. The azimuthal scan for (0,0,14) is also shown in the inset (black curve for 5 eV, red curve for 10 eV, multiplied by 5).
b) Re L$_3$-edge SS $(0,0,4n+2)$ energy scans. c) SP $(0,0,4n+2)$ energy scans with 1\% leakage from the SS channel. The inset is a zoom in to highlight the different energy positions of the lowest-energy peak for $(0,0,2)$. This and other interference features are discussed in the text in order to identify the presence of a magnetic component in the signal.}
\label{figSP1}
\end{figure}

In an ideal experiment, $f_{Q_0}$ does not contribute to SP scattering and the magnetic contribution $f_{m_z}$ can be easily separated from the Templeton term $f_{Q_{\tilde{x}^2-\tilde{y}^2}}$ because of their different azimuthal dependence, the magnetic term being proportional to $\sin^2(2\phi)$, the Templeton term (as seen above in SP geometry) to $\cos^2(2\phi)$. The zero of the azimuthal scan is always given relative to the cubic $\vec{a}$-axis.
Therefore, a signal measured at $45^\circ$ is purely due to magnetism, analogous to the case of Cd$_2$Os$_2$O$_7$ \cite{yamaura}. This is further confirmed by the numerical calculations performed with the FDMNES program
for the $I\overline{4}'m'2$ magnetic space group (relativistic calculation with two magnetic 5d electrons per Re ion and atomic positions of $I\overline{4}m2$ \cite{huang}) and shown in Fig.~\ref{figSP1}a. Its intensity is purely magnetic. 

However, in a real experiment, a leakage from the SS channel of around $1\%$ would contaminate this magnetic signal. The SS Bragg scattering is shown in Fig.~\ref{figSP1}b. Luckily, the magnetic term, usually out-of-phase by $\pi/2$ with the Bragg term, brings with it an extra $i$ coming from the $i \sin(2\pi u(4n+2))$ factor. Therefore, as already shown in Eq.~(\ref{strfacmag}), the magnetic contribution $f_{m_z}$ is in phase with the Bragg scattering of SS origin $f_{Q_0}$ (contrary to the quadrupolar terms $Q_{\tilde{x}^2-\tilde{y}^2}$, out of phase with both). This situation is shown in Fig.~\ref{figSP1}c, where the magnetic signal can still be clearly identified as a modulation of the SS signal. Three features can be experimentally identified to provide evidence of a magnetic signal in SP geometry, as a consequence of the interference with the Bragg scattering (we remind that in the case of no magnetism, the SP energy scan would be just a reduced intensity version of the SS energy scan): 1) the maximum of the $(002)$ SP reflection is shifted to a lower energy compared to the other peaks (inset of Fig.~\ref{figSP1}c). This condition implies interference between $f_{Q_0}$ and $f_{m_z}$ that is not present in the ideal case (Fig.~\ref{figSP1}a): in the case of a pure magnetic signal, all peaks are proportional to each other. 2) The peak at the $t_{2g}$ energy is bigger than that at the $e_g$ energy for $(0,0,10)$ and $(0,0,14)$ (Fig.~\ref{figSP1}c). This is not the case for the Bragg SS signal of Fig.~\ref{figSP1}b, having three peaks instead. 3) Finally, the tails of the Bragg signal are characterized by a Lorentzian decrease as a function of the energy whereas the magnetic peaks usually fall off faster than a Lorentzian.
In any case, the above signal must be measured at $45^\circ$ (or at $135^\circ$), where $Q_{\tilde{x}^2-\tilde{y}^2}$ is zero.

To conclude this subsection on the search for magnetic ordering at $T_{c1}$, we remark that it is in principle also possible to look for a magnetic signal determined by the magnetic quadrupole (and not by the dipole component $m_z$, as above). This can be done at the L$_1$ pre-edge by E1-E2 transitions, that are allowed by inversion-breaking. Several multipoles can contribute in principle to the pre-edge intensity (the full list can be found, e.g., in Ref.~\onlinecite{sergioxncd}). For the $I\overline{4}'m'2$ magnetic group, only the magnetic quadrupole and the axial toroidal quadrupoles can be detected, so that if we could single out the two signals, their dependence as a function of $T_{c1}-T$ is expected to follow a square root and a linear dependence, respectively, in keeping with the SHG measurement. However, the interest in this approach is unfortunately strongly reduced by our FDMNES calculations, showing, in the same energy range as the E1-E2 terms, the presence of E2-E2 contributions (mainly determined by an hexadecapolar OP), which are 10 to 30 times bigger, and would completely hide the inversion-odd signal of E1-E2 origin.

\subsection{Probing by REXS the 120 K phase transition of ${\rm Cd_2Re_2O_7}$}

Two different phase transitions have been reported in Cd$_2$Re$_2$O$_7$. Besides the phase transition at $T_{c1}$, from the high-temperature $Fd\overline{3}m$ cubic phase to the $I\overline{4}m2$ phase, studied above, a second phase transition has been measured around $T_{c2}\simeq 120$ K \cite{hiroi,huang}. The crystal space group below $T_{c2}$ was identified as $I4_122$ \cite{huang}. However, several doubts were raised in the literature about its existence \cite{disorder}. Here we propose a REXS experiment that can definitely settle the issue.

\begin{table}[ht!]
	\caption{Fractional positions of Re atoms in the $I4_122$ space group \cite{huang}. The four bcc translations are not shown.  Of all symmetry operations relative to Re$_1$, only the ones used to derive the resonant structure factor are highlighted.}
	\centering
	\begin{ruledtabular}
		\begin{tabular}{c|c|c}
			Atom & (x,y,z)~$I4_122$ & $I4_122$ \\
			\colrule
			Re$_1$ & (0.9967,0.25,0.125)   & ${\hat{E}}$   \\
			Re$_2$ & (0.5033,0.25,0.625) & ${\hat{C}}_{2z}$   \\
			Re$_3$ & (0.75,0.4967,0.375) & ${\hat{C}}_{4z}$  \\
			Re$_4$ & (0.75,0.0033,0.875) & ${\hat{C}}_{4z}^-$  \\
		\end{tabular}
	\end{ruledtabular}
	\label{sym2}
\end{table} 

The resonant X-ray structure factor for the $I4_122$ space group, summed over the 8 Re atoms, can be written as: 

\begin{align} \label{strfac2}
& F^{hkl}_{I4_122} \propto  (1+(-1)^{h+k+l})  \\
 & \hspace{0.25in} (e^{2i\pi (hs+k/4+l/8)}(1+(-)^{h+l} e^{-4i\pi hs}{\hat{C}}_{2z})+ \nonumber \\
 & \hspace{0.25in} e^{2i\pi (3h/4+ks+k/2+3l/8)}(1+(-)^{k+l} e^{-4i\pi ks}{\hat{C}}_{2z}){\hat{C}}_{4z})f_1  \nonumber
\end{align}
Here, $s=0.9967$ \cite{huang}.

\begin{figure}[ht!]
\includegraphics[width=0.45\textwidth]{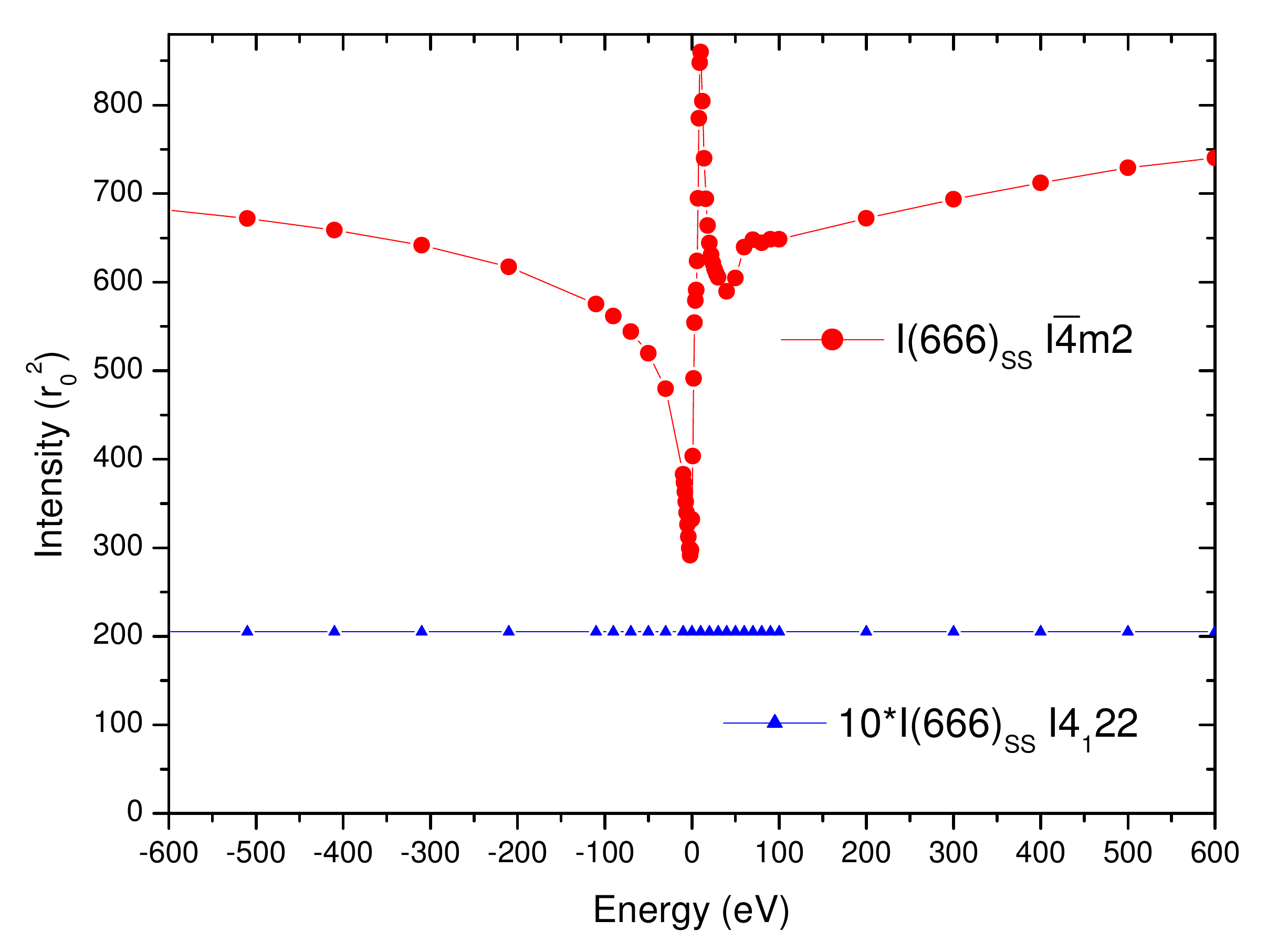}
\caption {The REXS intensity for the (666) reflection near the Re L$_3$ edge. The black curve is calculated by FDMNES for the $I4_122$ space group and is multiplied by 10. The weak intensity, constant in energy, is due to the non-resonant contribution from the oxygens. The red curve represents the calculation for the ${I\overline{4}m2}$ space group. The strong energy dependence is clearly seen at the Re L$_3$ edge. Far from the edge, the signal is about 35 times bigger than for $I4_122$. Both curves are calculated in SS geometry.}
\label{rexs}
\end{figure}

In order to have a clear differentiation of the $I4_122$ and ${I\overline{4}m2}$  space groups, we need to look for reflections of the kind $(h,h,4n+2)$ near the resonant energy of Re ions. In fact, these reflections are forbidden at the special positions ($8f$) of Re ions in the $I4_122$ space group (ITC No.~98) for SS geometry. The only contribution to it would come from oxygen sites O1 ($8d$) and O3 ($8e$), therefore non-resonant. 
 Instead, for the ${I\overline{4}m2}$ space group, a clear resonant behavior appears in the SS geometry. A specific calculation by FDMNES for the (666) reflection near the Re L$_3$-edge shows this quantitatively, as seen in Fig.~\ref{rexs}. Therefore a resonant behavior onsetting at $T_{c1}$ is the signature of the ${I\overline{4}m2}$ space group. A change to the non-resonant behavior at $T_{c2}$ would signal the transition to $I4_122$. 

\section{Discussion and conclusions}

As we demonstrated above, a magnetic scenario gives a natural interpretation for the primary even-parity and odd-parity order parameters implied by the second harmonic generation data. This is in contrast to a structural scenario, where only the interpretation of the secondary order parameter as an axial toroidal quadrupole, common to the magnetic interpretation, is obvious.
On the other hand, there is no evidence for magnetic order from NMR and NQR data. Such order should show up as a splitting of the lines, which to date has not been observed \cite{noteOs}. Both our magnetic patterns should show such a splitting, unless the actual magnetic order parameter at the Re site itself is a magnetic octupole. However, both polarized REXS and polarized neutrons should be able to settle the question of magnetic order in the Re compound. In the case of REXS, we have explicitly shown how, in Section IV. Regardless, we think that a magnetic quadrupole/octupole
explanation for the primary odd/even-parity order parameters is an attractive possibility worth exploring. 

The other issue concerns the fact that Harter {\it et al.}'s SHG data are most consistent with the $\epsilon^{\rm out}_{z'}$ component being in quadrature with $\epsilon^{\rm out}_{x'}$ and $\epsilon^{\rm out}_{y'}$ (here, outgoing polarization components refer to the last line of Eq.~(\ref{polax})). From our exact quantum mechanical expressions, these quantities should instead interfere. 
As we stated in Section II.B and in Appendix B, a possible solution is represented by either birefringence or refraction. In the latter case, we suppose that the effective $\theta$ inside the sample could be lower because of refraction, thus suppressing the interference with the $\epsilon^{\rm out}_{z'}$ term. This could be tested by increasing the incidence angle of the incoming beam as shown in Fig.~\ref{fignew}. As discussed at the end of Section II.B, this case could also allow a clear identification of the magnetic OP, whether $M_{3z^2-r^2}$ or $M_{xy}$, because of their different interference properties.  
Birefringence was already proposed to explain outgoing elliptical polarized radiation in Petersen {\it et al.}~\cite{petersen}, but the geometry of this SHG experiment had $\epsilon^{\rm out}_{z'}=0$ and so it might not be directly related to the case of Harter {\it et al.}~\cite{harter}. 

\begin{figure}[ht!]
\includegraphics[width=0.4\textwidth]{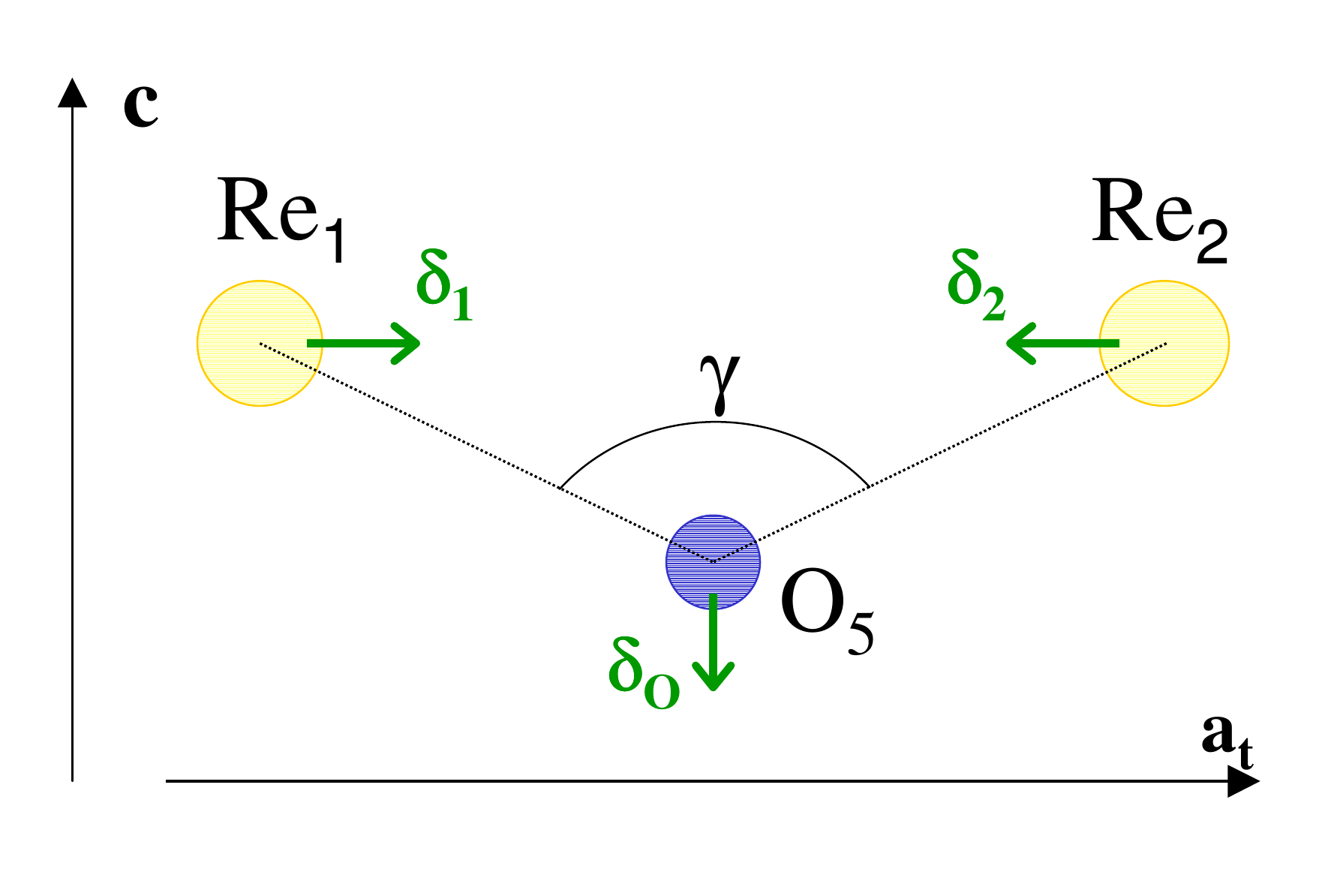}
\caption {Pictorial representation of the Re$_1$-O$_5$-Re$_2$ distortion that would increase the DM interaction (see text).}
\label{dmi}
\end{figure}

Also in this paper, we have avoided any real discussion concerning microscopics, but we offer a few remarks here.  First, unlike the
Os compound, which is $5d^3$ and an insulator, the Re compound is $5d^2$ and a metal.  Surely, these two differences have something to do with the differences between the two materials, which should be investigated by further theoretical studies beyond those done in
Ref.~\onlinecite{huang}.  And, in regards to the spin nematic scenario of Ref.~\onlinecite{fu}, we wish to point out that from the work
of Ref.~\onlinecite{huang}, the atomic spin-orbit coupling is already large enough to cause a large splitting of the originally Kramers
degenerate Fermi surfaces (i.e., their Figure 8).  So, why an additional splitting of a similar form would occur from another mechanism is not apparent to us.

We also note that the two magnetic structures proposed in Figs.~\ref{magpat}(a) and \ref{magpat}(b) correspond, respectively, to particular cases of the so-called indirect and direct magnetic configurations \cite{lacroix} induced by the Dzyaloshinsky-Moriya (DM) interaction in a pyrochlore lattice. This line of thought is further motivated by the recent study of the microscopic origin of the DM interaction in $d^1$ pyrochlores \cite{arakawa}, where it is shown that it arises from a bond corresponding, in our case, to Re$_1$-O$_5$-Re$_2$ in Fig.~\ref{pyrochlore}. Although this bond locally breaks inversion symmetry, it keeps the mirror symmetry in the Re$_1$-O$_5$-Re$_2$ plane associated with time-reversal symmetry (i.e., the $m'$ in the $I\overline{4}'m'2$ space group). We show this in Fig.~\ref{dmi} in the $ac$ tetragonal plane. The strength of the DM interaction, $D$, is determined by this bond angle (Fig.~2 of Ref.~\onlinecite{arakawa}). Its value increases with decreasing $\gamma$ in Fig.~\ref{dmi} ($D$ being zero when this angle is $180^\circ$). This is exactly the trend at the $T_{c1}$ transition, where, as shown in Fig.~\ref{dmi}, in the tetragonal phase Re$_1$ and Re$_2$ move towards each other, whereas O$_5$ moves down by $\vec{\delta}_O$. Thus, the bond angle decreases to $136.1^\circ$ (versus a value of $139.6^\circ$ in the cubic phase), thereby increasing $D$ \cite{notefin}. We remark that this is the largest Re-O-Re angle change at the transition and might explain the structural coupling to the magnetic degrees of freedom in Cd$_2$Re$_2$O$_7$: the angle tends to decrease, so as to increase the DM strength $D$, at the expense of elastic energy, and a new minimum is found with a tetragonal distortion. In this picture, the magnetic degrees of freedom would be primary, the distortion a secondary consequence.  
Whether the scenario of Ref.~\onlinecite{arakawa} works as well in $5d^2$ pyrochlores like Cd$_2$Re$_2$O$_7$ is however beyond the scope of the present paper.

Finally, we would like to emphasize that SHG is a wonderful method to identify hidden order associated with novel electronic states and magnetic configurations characterized by higher order, parity-odd, multipoles (e.g., magnetic quadrupoles). We remark that such multipoles are not necessarily associated with orbital currents.  In the future, we expect that a photon energy sweep will be critical in helping to unravel the nature of the optical transitions, in particular the electronic states involved in the transition, so as to provide further insight into the origin of the observed SHG signals in Cd$_2$Re$_2$O$_7$ and Sr$_2$IrO$_4$. To this aim, we also believe that further advances in understanding can be obtained by developing a full polarization analysis of both linear and circular polarization channels: having access to circular polarization, for example, would be useful to investigate the origin of the $\Phi\sim\pi/2$ phase shift noted by Petersen {\it et al.}~\cite{petersen} and discussed in Appendix B. We remind that linear incoming polarization can be transformed to circular outgoing polarization by two resonances that are close in energy \cite{mazzoli}.

\begin{acknowledgments}
The authors would like to thank Dr.~John Harter and Prof.~David Hsieh for discussions concerning their data.
Work by MRN was supported by the Materials Sciences and Engineering
Division, Basic Energy Sciences, Office of Science, US DOE.
\end{acknowledgments}

\appendix

\section{Explicit multipolar expression of SHG amplitudes}

We give here the explicit general expressions for the dipole and octupole terms of the E1-E1-E1 transitions, not given in Section II. The two dipoles are (here ${\alpha}$ is any of $x$, $y$ or $z$):

\begin{align}
\bar{O}_{\alpha}^{(1)} = \frac{1}{\sqrt{3}}\epsilon_{\alpha}^o \vec{\epsilon}^i\cdot \vec{\epsilon}^i \leftrightarrow \big(\chi_{\alpha xx}+\chi_{\alpha yy}+\chi_{\alpha zz}\big)/\sqrt{3} ;
\label{tordip1cart}
\end{align}

\begin{align}
\tilde{O}_{\alpha}^{(2)} = & \frac{1}{\sqrt{15}} {\epsilon}^{o}_{\alpha} \vec{\epsilon}^i\cdot \vec{\epsilon}^i - \frac{3}{\sqrt{15}} \epsilon_{\alpha}^i \vec{\epsilon}^o \cdot \vec{\epsilon}^i  \leftrightarrow \nonumber \\ 
& \frac{1}{\sqrt{15}} \big[\chi_{\alpha xx}+\chi_{\alpha yy}+\chi_{\alpha zz}-\frac{3}{2}(\chi_{x\alpha x}+\chi_{xx\alpha} \nonumber \\
&\,\,\,\,\,\,\,\,\,+\chi_{y\alpha y}+\chi_{yy\alpha}+\chi_{zz\alpha}+\chi_{z\alpha z})\big]
\label{tordip2cart}
\end{align}

We remind that the dipoles form $T_{1u}$ irreps in $O_h$ symmetry.

The octupole terms, in the axial representation of DMN \cite{prbIr}, can be written as:
\begin{align} \label{toroctcart}
& \tilde{O}_{y(3x^2-y^2)}  = \frac{1}{2} \big[ {\epsilon}^{o}_y (\epsilon^i_x\epsilon^i_x - \epsilon^i_y\epsilon^i_y ) +2 {\epsilon}^{o}_x \epsilon^i_x \epsilon^i_y \big] \\
& \tilde{O}_{x(3y^2-x^2)} = \frac{1}{2} \big[ {\epsilon}^{o}_x (\epsilon^i_y\epsilon^i_y - \epsilon^i_x\epsilon^i_x ) +2 {\epsilon}^{o}_y \epsilon^i_y \epsilon^i_x \big]  \nonumber \\
& \tilde{O}_{z(x^2-y^2)} = \sqrt{\frac{2}{3}} \big[\epsilon^i_z ({\epsilon}^{o}_x\epsilon^i_x - {\epsilon}^{o}_y\epsilon^i_y) + \frac{1}{2}{\epsilon}^{o}_z (\epsilon^i_x\epsilon^i_x - \epsilon^i_y\epsilon^i_y )  \big] \nonumber \\
& \tilde{O}_{xyz}  = \sqrt{\frac{2}{3}} \big[ {\epsilon}^{o}_z \epsilon^i_x\epsilon^i_y + {\epsilon}^{o}_x \epsilon^i_z\epsilon^i_y + {\epsilon}^{o}_y \epsilon^i_x\epsilon^i_z  \big]  \nonumber \\
& \tilde{O}_{z^3} = \sqrt{\frac{2}{45}} \big[ (5{\epsilon}^{o}_z \epsilon^i_z-3\vec{{\epsilon}}^{o} \cdot \vec{\epsilon^i})\epsilon^i_z + \frac{1}{2}\epsilon^o_z (5{\epsilon}^{i}_z \epsilon^i_z-3\vec{{\epsilon}}^{i} \cdot \vec{\epsilon^i}) \big] \nonumber \\
& \tilde{O}_{xz^2} = \frac{1}{\sqrt{15}} \big[ (5{\epsilon}^{o}_z \epsilon^i_z-\vec{{\epsilon}}^{o} \cdot \vec{\epsilon^i})\epsilon^i_x + \frac{1}{2}\epsilon^o_x (5{\epsilon}^{i}_z \epsilon^i_z-\vec{{\epsilon}}^{i} \cdot \vec{\epsilon^i}) \big] \nonumber \\
& \tilde{O}_{yz^2} = \frac{1}{\sqrt{15}} \big[ (5{\epsilon}^{o}_z \epsilon^i_z-\vec{{\epsilon}}^{o} \cdot \vec{\epsilon^i})\epsilon^i_y + \frac{1}{2}\epsilon^o_y (5{\epsilon}^{i}_z \epsilon^i_z-\vec{{\epsilon}}^{i} \cdot \vec{\epsilon^i}) \big]  \nonumber 
\end{align}

We remark that we have rewritten this equation in a slightly different form than in Eq.~(A48) of DMN \cite{prbIr}, in order to highlight the symmetry of the tensor, that was less evident in the original form \cite{prbIr}. 
In fact, we remind that $\tilde{O}_{z^3}$ is a shorthand notation for $\tilde{O}_{z(5z^2-3r^2)}$, $\tilde{O}_{xz^2}$ for $\tilde{O}_{x(5z^2-r^2)}$ and $\tilde{O}_{yz^2}$ for $\tilde{O}_{y(5z^2-r^2)}$.
Eq.~(\ref{toroctcart}) is however equal, line by line, to the octupole expression in Eq.~(A48) of DMN. 

When the symmetry is cubic, it is more appropriate to work with the irreps of the octupole under $O_h$: $T_{1u}$ and $T_{2u}$ triplets and the singlet, $A_{2u}$. We give the corresponding $\chi_{ijk}$ irreducible components in this case.

The $T_{1u}$ triplet can be obtained by combining the first two and the last three terms of Eq.~(\ref{toroctcart}):

\begin{align}
& \tilde{O}_{x^3} = \sqrt{\frac{2}{45}} \big[ (5{\epsilon}^{o}_x \epsilon^i_x-3\vec{{\epsilon}}^{o} \cdot \vec{\epsilon^i})\epsilon^i_x + \frac{1}{2}\epsilon^o_x (5{\epsilon}^{i}_x \epsilon^i_x-3\vec{{\epsilon}}^{i} \cdot \vec{\epsilon^i}) \big]  \leftrightarrow  \nonumber \\
& \sqrt{\frac{2}{5}}\big[\chi_{xxx}-\frac{\chi_{yyx}+\chi_{yxy}+\chi_{xyy}}{2}-\frac{\chi_{zzx}+\chi_{zxz}+\chi_{xzz}}{2}\big]; \nonumber \\ 
& \tilde{O}_{y^3} = \sqrt{\frac{2}{45}} \big[ (5{\epsilon}^{o}_y \epsilon^i_y-3\vec{{\epsilon}}^{o} \cdot \vec{\epsilon^i})\epsilon^i_y + \frac{1}{2}\epsilon^o_y (5{\epsilon}^{i}_y \epsilon^i_y-3\vec{{\epsilon}}^{i} \cdot \vec{\epsilon^i}) \big]  \leftrightarrow \nonumber \\
& \sqrt{\frac{2}{5}}\big[\chi_{yyy}-\frac{\chi_{xxy}+\chi_{xyx}+\chi_{yxx}}{2}-\frac{\chi_{zzy}+\chi_{zyz}+\chi_{yzz}}{2}\big]; \nonumber \\
& \tilde{O}_{z^3} = \sqrt{\frac{2}{45}} \big[ (5{\epsilon}^{o}_z \epsilon^i_z-3\vec{{\epsilon}}^{o} \cdot \vec{\epsilon^i})\epsilon^i_z + \frac{1}{2}\epsilon^o_z (5{\epsilon}^{i}_z \epsilon^i_z-3\vec{{\epsilon}}^{i} \cdot \vec{\epsilon^i}) \big] \leftrightarrow \nonumber \\
& \sqrt{\frac{2}{5}}\big[\chi_{zzz}-\frac{\chi_{xxz}+\chi_{xzx}+\chi_{zxx}}{2}-\frac{\chi_{yyz}+\chi_{yzy}+\chi_{zyy}}{2}\big]. 
\label{T1uoct}
\end{align}
where $\tilde{O}_{x^3}$ is a shorthand notation for $\tilde{O}_{x(5x^2-3r^2)}$, and $\tilde{O}_{y^3}$ for $\tilde{O}_{y(5y^2-3r^2)}$.

The $T_{2u}$ triplet can be obtained by combining the first three and the last two terms of Eq.~(\ref{toroctcart}):

\begin{align}
& \tilde{O}_{x(y^2-z^2)} = \frac{1}{\sqrt{6}} \big[ 2\epsilon^i_x ({\epsilon}^{o}_y\epsilon^i_y - {\epsilon}^{o}_z\epsilon^i_z) + {\epsilon}^{o}_x (\epsilon^i_y\epsilon^i_y - \epsilon^i_z\epsilon^i_z )  \big] \leftrightarrow  \nonumber \\
& (\chi_{xyy}-\chi_{xzz}+\chi_{yxy}+\chi_{yyx}-\chi_{zxz}-\chi_{zzx})/\sqrt{6} ; \nonumber \\ 
& \tilde{O}_{y(z^2-x^2)} = \frac{1}{\sqrt{6}} \big[ 2\epsilon^i_y ({\epsilon}^{o}_z\epsilon^i_z - {\epsilon}^{o}_x\epsilon^i_x) + {\epsilon}^{o}_y (\epsilon^i_z\epsilon^i_z - \epsilon^i_x\epsilon^i_x )  \big] \leftrightarrow \nonumber \\
& (\chi_{yzz}-\chi_{yxx}+\chi_{zyz}+\chi_{zzy}-\chi_{xyx}-\chi_{xxy})/\sqrt{6}  ; \nonumber \\
& \tilde{O}_{z(x^2-y^2)} = \frac{1}{\sqrt{6}} \big[ 2\epsilon^i_z ({\epsilon}^{o}_x\epsilon^i_x - {\epsilon}^{o}_y\epsilon^i_y) + {\epsilon}^{o}_z (\epsilon^i_x\epsilon^i_x - \epsilon^i_y\epsilon^i_y )  \big] \leftrightarrow \nonumber \\
& (\chi_{zxx}-\chi_{zyy}+\chi_{xzx}+\chi_{xxz}-\chi_{yzy}-\chi_{yyz})/\sqrt{6} ; 
\label{T2uoct}
\end{align}

Finally, the $A_{2u}$ singlet corresponds to the fourth term of Eq.~(\ref{toroctcart}):

\begin{align} \label{A2uoct}
\tilde{O}_{xyz} = \sqrt{\frac{2}{3}} \big[ {\epsilon}^{o}_z \epsilon^i_x\epsilon^i_y + {\epsilon}^{o}_x \epsilon^i_z\epsilon^i_y + {\epsilon}^{o}_y \epsilon^i_x\epsilon^i_z  \big] \leftrightarrow \nonumber \\
(\chi_{xyz}+\chi_{xzy}+\chi_{yzx}+\chi_{yxz}+\chi_{zxy}+\chi_{zyx})/\sqrt{6}.
\end{align}

Eqs.~(\ref{tordip1cart}), (\ref{tordip2cart}), (\ref{T1uoct}), (\ref{T2uoct}) and (\ref{A2uoct}), together with Eq.~(\ref{magquadcart}) in Section II.A, represent the recoupling scheme in spherical tensors of the SHG amplitude, valid in the general case for $O_h$ and its subgroups. In the case of axial symmetries (e.g., $C_6$), the first two terms of Eq.~(\ref{T1uoct}) and the first two terms of Eq.~(\ref{T2uoct}) should be replaced by the first two and the last two terms of Eq.~(\ref{toroctcart}).

\section{Calculation of the SHG azimuthal scans for the allowed susceptibilities of the $I\overline{4}m2$ subgroups}

\begin{figure}[ht!]
\includegraphics[width=0.4\textwidth]{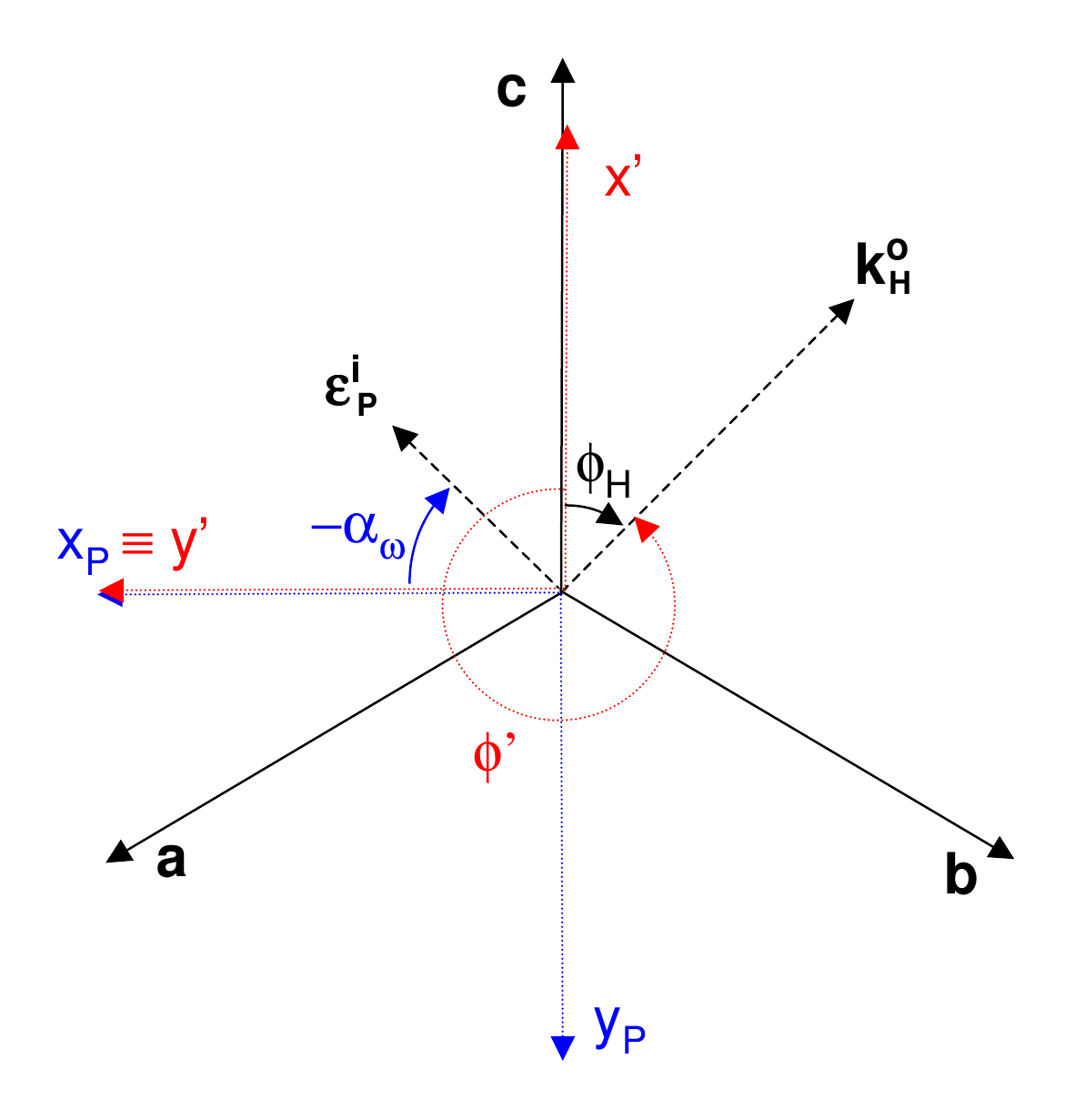}
\caption{Projection of the SHG experiment onto the plane perpendicular to the (111) cubic direction. The $z'$ axis, parallel to the (111) cubic axis, is out of the plane of the drawing. Both $\phi'$ and Harter {\it et al.}'s $\phi_H$ refer to the angle of the projection of
Harter {\it et al.}'s $\vec{k}^o_H$ in the $x'$-$y'$ plane. Petersen {\it et al.}'s $\alpha_{\omega}$ refers instead to the incoming in-plane polarization vector, as $\vec{k}^o_P$ is directed along the (111) cubic direction.}
\label{azhar}
\end{figure}

The azimuthal scan in Harter {\it et al.}~\cite{harter} was performed around the (111) cubic direction. Here we calculate the transformation from the coordinate set associated with the cubic $\vec{a}$, $\vec{b}$, $\vec{c}$ axes of the high-temperature $Fd\overline{3}m$ space group (called $x$, $y$, $z$, here and in the rest of the paper) and the coordinate set associated with the azimuthal scan, called $x'$, $y'$, $z'$, with the $z'$ axis along the (111) cubic direction and the $x'$ axis parallel to the projection of the $\vec{c}$ axis in the (111) plane, as shown in Fig.~\ref{azhar}. We have:

\begin{align}
x'= & \frac{1}{\sqrt{6}}(-x - y + 2z) \nonumber \\
y'= & \frac{1}{\sqrt{2}}(x - y)  \nonumber \\
z'= & \frac{1}{\sqrt{3}}(x + y + z) 
\label{cubax}
\end{align}

The opposite transformations can be written as: 

\begin{align}
x= & -\frac{x'}{\sqrt{6}} + \frac{y'}{\sqrt{2}} + \frac{z'}{\sqrt{3}} \nonumber \\
y= & -\frac{x'}{\sqrt{6}} - \frac{y'}{\sqrt{2}} + \frac{z'}{\sqrt{3}} \nonumber \\
z= & \sqrt{\frac{2}{3}}x' + \frac{z'}{\sqrt{3}}
\label{axcub}
\end{align}

We remark that the usual azimuthal angle $\phi'$ goes from $x'$ to $y'$ and has the opposite rotation as the azimuthal angle chosen by Harter {\it et al.}~\cite{harter}: $\phi_H= - \phi'$. Petersen {\it et al.}'s angle, $\alpha$, is related to the 90$^\circ$-shifted $x_P-y_P$ frame, but it refers to the $\epsilon_S$ component, as shown in Fig.~\ref{azhar}. 
The incoming and outgoing S and P polarizations in the $x'$, $y'$, $z'$ frame of the azimuthal scan for Harter {\it et al.} experiment are given by: 

\begin{align}
& \vec{\epsilon}_S^{in} =  (\sin\phi_H,\cos\phi_H,0)=\vec{\epsilon}_S^{out} ; \nonumber \\
& \vec{\epsilon}_P^{in} =  (\cos\theta \cos\phi_H,-\cos\theta \sin\phi_H,\sin\theta) ; \nonumber \\
& \vec{\epsilon}_P^{out} =  (-\cos\theta \cos\phi_H,\cos\theta \sin\phi_H,\sin\theta)
\label{polax}
\end{align}

The angle $\theta$ is the angle between the incident beam and the (111) direction.

We remind that in Section II and III, tensor labels $\chi_{ijk}$ have been given with respect to the cubic crystal axes, even in the low-temperature $I\overline{4}m2$ phase. So, it is necessary to transform incoming and outgoing polarizations, associated with the non-zero susceptibilities, in the cubic frame. 
We get: 

\begin{align}
\vec{\epsilon}_S^{in} & =\vec{\epsilon}_S^{out}  =  (-\sin\phi_H/\sqrt{6}+\cos\phi_H/\sqrt{2},  \nonumber \\
& -\sin\phi_H/\sqrt{6}-\cos\phi_H/\sqrt{2},\sqrt{2/3}\sin\phi_H) ; \nonumber \\
\vec{\epsilon}_P^{in} & =  (-\cos\theta\cos\phi_H/\sqrt{6}-\cos\theta\sin\phi_H/\sqrt{2}+\sin\theta/\sqrt{3}, \nonumber \\
& -\cos\theta\cos\phi_H/\sqrt{6}+\cos\theta\sin\phi_H/\sqrt{2}+\sin\theta/\sqrt{3},  \nonumber \\
& +\sqrt{2/3}\cos\theta\cos\phi_H+\sin\theta/\sqrt{3}) ; \nonumber \\
\vec{\epsilon}_P^{out} & =  (\cos\theta\cos\phi_H/\sqrt{6}+\cos\theta\sin\phi_H/\sqrt{2}+\sin\theta/\sqrt{3},  \nonumber \\
& \cos\theta\cos\phi_H/\sqrt{6}-\cos\theta\sin\phi_H/\sqrt{2}+\sin\theta/\sqrt{3},  \nonumber \\
& -\sqrt{2/3}\cos\theta\cos\phi_H+\sin\theta/\sqrt{3})
\label{polcub}
\end{align}

Using the definition of the quadrupole and octupole polarization tensors given in Eqs.~(\ref{magquadcart}), (\ref{T2uoct}) and (\ref{A2uoct}), we get the azimuthal contributions for each of the relevant tensors in the $I\overline{4}m2$ subgroups that are reported in Section II.B, Eq.~(\ref{azscanphi}). 

It is then possible to reproduce the experimental data of Fig.~4a of Petersen {\it et al.}~\cite{petersen} with only the second line of Eq.~(\ref{azscanphi}), i.e., the term $\tilde{O}_{x^2-y^2}$ associated with the axial toroidal quadrupole $G_{x^2-y^2}$. In fact, the experimental geometry is in this case limited to both incoming and outgoing polarizations, $\vec{\epsilon}^{i}$ and $\vec{\epsilon}^{o}$, lying in the (111) plane (we remind that $\theta=0^\circ$ in Ref.~\onlinecite{petersen}). The two polarizations are associated, respectively, with the angles $\alpha_{\omega}$ and $\alpha_{2\omega}$, both varying from $0$ to $\pi$. This is a mixed configuration between SS and SP. In fact, the outgoing in-plane electric polarization can have both a component along the incoming electric polarization (which would give an SS configuration with zero intensity) and a component perpendicular to the incoming polarization (leading to a signal due to the SP configuration). The non-zero projection in the SP channel is proportional to $\sin|\alpha_{\omega}-\alpha_{2\omega}|$. This must be weighted by the rotation of the incoming polarization vector from $\alpha_{\omega}=0$ (parallel to $x_P$) to $\alpha_{\omega}=\pi$ (antiparallel to $-x_P$), leading to a factor $\cos\alpha_{\omega}$, from the $\tilde{O}_{x^2-y^2}$ term of the second line of Eq.~(\ref{azscanphi}) with $\theta=0$. Squaring the result for the intensity, we get $\cos^2\alpha_{\omega} \sin^2(\alpha_{\omega}-\alpha_{2\omega})$, as reported in Section II.B and in Fig.~\ref{figpet}(a), left plot. In the case of $I4_122$, we would have gotten instead the contribution of the first term of Eq.~(\ref{azscanphi}), $\tilde{O}_{3z^2-r^2}$, this time $\sin^2\alpha_{\omega}$ (again for $\theta=0$). So the final result is: $\sin^2\alpha_{\omega} \sin^2(\alpha_{\omega}-\alpha_{2\omega})$, as plotted in Fig.~\ref{figpet}(a), right plot. We remark that a proper description of the experimental data, as in the original paper \cite{petersen}, cannot be obtained in this way, and an extra parameter $\Phi$ has to be added to describe the effect of birefringence. 
The value quoted in Petersen {\it et al.}~is $\Phi=1.1 e^{1.3i}$. We remark that the effect of such a parameter is to add a phase shift of roughly $\pi/2$ between the outgoing polarizations in the $x'$ and $y'$ directions. This means that the outgoing electric field is elliptically (and almost circularly) polarized.

In order to reproduce the results of Harter {\it et al.}~\cite{harter}, with $\theta=10^\circ$ ($\cos\theta\sim 0.985$, $\sin\theta\sim 0.174$), a second component is needed going like $\sin\phi_H$, either from $\tilde{O}_{3z^2-r^2}$ or from $\tilde{O}_{xy}$, as discussed in the text. It is then possible to reproduce the experimental data by fitting the second and the third (or the first) line of Eq.~(\ref{azscanphi}) together with a surface component going like $\cos(3\phi_H)$. This is shown in Fig.~\ref{figpet}(b). In particular, the green dashed lines in Fig.~\ref{figpet}(b) is obtained through $\tilde{O}_{3z^2-r^2}$, $\tilde{O}_{x^2-y^2}$ and the surface component, whereas the red dashed line is a linear combination of $\tilde{O}_{xy}$, $\tilde{O}_{x^2-y^2}$ and the surface component.  
 
However, again, a proper description of the experimental data cannot be obtained, mainly because of the interference between the $\cos\phi_H$ and the $\cos(2\phi_H)$ terms in $\tilde{O}_{x^2-y^2}$. This is visible in the deformed shape of the theoretical plot in the right frame of Fig.~\ref{figpet}(b) for $\phi_H=0$ and $\phi_H=\pi$. Such a discrepancy might be explained in terms of a lower effective $\theta$ angle for the incoming beam, due to refraction, that would reduce the interference with $\cos(2\phi_H)$ in $\tilde{O}_{x^2-y^2}$, as the latter term is weighted by $\sin\theta$ \cite{harternote}.
If this were the case, we might expect to resolve the issue about whether the primary OP is determined by $\tilde{O}_{3z^2-r^2}$ or by $\tilde{O}_{xy}$, by exploiting the different phase of the interference ($\sin\phi_H$ versus $\sin(2\phi_H)$) in the two terms (opposite sign for $\tilde{O}_{xy}$, same sign for $\tilde{O}_{3z^2-r^2}$), as described at the end of Section II.B and shown in Fig.~\ref{fignew}(c).

We conclude with a remark on the tetragonal set of coordinates for the space group $I\overline{4}m2$. They correspond to a rotation of the cubic coordinates by $45^\circ$ around the $\vec{c}$ axis. For such a set, say $\tilde{x}$, $\tilde{y}$, $z$, the mirror symmetry becomes perpendicular to the $\tilde{x}$ or $\tilde{y}$ axes (it was $45^\circ$ from the $x$ and $y$ in the cubic frame), and the two-fold rotation is along the diagonals in the $\tilde{x}\tilde{y}$ plane (it was along $x$ and $y$ axes in the cubic frame), as can be deduced from Fig.~\ref{pyrochlore}.
This implies a label switching of the susceptibilities, in the same way as used for the electric quadrupole operator in Section IV: $\chi_{xyz} + \chi_{yxz} = \chi_{\tilde{x}\tilde{x}z} - \chi_{\tilde{y}\tilde{y}z}$. This is valid for any point group. 
In particular, for point group $\overline{4}2m$ with the two-fold rotations along $x$ and $y$, we have $\chi_{xyz} = \chi_{yxz}$. This is what is used throughout the whole paper (tetragonal space group described with cubic axes) with the alternative notation $\overline{4}m2$, for reasons explained at the beginning of Section II.B.
Actually, the $\overline{4}m2$ notation describes the same physical situation when expressed in the $45^\circ$-rotated $\tilde{x}\tilde{y}$ frame with the two-fold rotations along the diagonals (we remind that in the ITC notation, the first symmetry operation after the $\overline{4}$-fold axis refers to the orthogonal coordinate axis, the second to the diagonal). In this tetragonal frame, the former relation becomes $\chi_{\tilde{x}\tilde{x}z} = - \chi_{\tilde{y}\tilde{y}z}$. So, as $\chi_{xyz} = \chi_{\tilde{x}\tilde{x}z}$, the same physical susceptibility appears with two different labels according to the coordinate choice in the $xy$ ($\tilde{x}\tilde{y}$) plane.  So, particular care should be taken in specifying the coordinate set associated with the susceptibilities.

\section{Spherical harmonics rotation from the local to the cubic basis}

In Section IV, we have determined that reflections of the kind $(0,0,4n+2)$ are sensitive to the electric-quadrupole operator $Q_{xy}$, with the labels referring to the cubic axes. However, the local oxygen-octahedral environment of the Re ions in the cubic frame is rotated with respect to the cubic crystallographic axes (and trigonally compressed). In order to evaluate the weight of Re $5d$ orbitals contributing to the signal, we should perform the rotation.
For Re$_1$ at (000), as shown (projected) in Fig.~\ref{pyrochlore}, the six nearest-neighbor oxygens ($48f$ Wyckoff label) are at positions: 
$O_1$=(-0.0652, 0.125, 0.125), $O_2$=(0.0652, -0.125, -0.125), $O_3$=(0.125, -0.0652, 0.125), $O_4$=(-0.125, 0.0652, -0.125), $O_5$=(0.125, 0.125, -0.0652), $O_6$=(-0.125, -0.125, 0.0652) \cite{huang}.

Referring to Fig~\ref{pyrochlore}, if we choose to orient the $z''$ axis from Re$_1$ to O$_6$, the $x''$ axis from Re$_1$ to O$_3$ and the $y''$ axis from Re$_1$ to O$_1$, then Euler angles are $\alpha=-45^\circ$, $\beta=69.75^\circ$, $\gamma = 45^\circ$. From the general transformation for spherical tensors ($\theta$ and $\phi$ are the usual spherical angles referred to the cubic frame, whereas $\theta''$ and $\phi''$ refer to the local frame with $z''$-axis along Re$_1$-O$_6$):

\begin{align}
&Q_{xy}(\theta,\phi)=-\frac{i}{\sqrt{2}}[Y_{2,2}(\theta,\phi)-Y_{2,-2}(\theta,\phi)] = \nonumber \\
&+Q_{y''z''}(\theta'',\phi'') \left(\sin\alpha\sin\beta\cos\beta\sin\gamma - \cos\alpha \sin\beta \cos\gamma \right) \nonumber \\
&+Q_{x''z''}(\theta'',\phi'') \left(\cos\alpha\sin\beta\cos\beta\sin\gamma + \sin\alpha \sin\beta \cos\gamma \right) \nonumber \\
&+Q_{x''y''}(\theta'',\phi'') [\cos(2\alpha) \cos\beta \cos\gamma \nonumber \\
& - \sin(2\alpha) \frac{1+\cos^2\beta}{2} \sin\gamma] \nonumber \\
& -Q_{x''^2-y''^2}(\theta'',\phi'') [\cos(2\alpha) \frac{1+\cos^2\beta}{2} \sin\gamma  \nonumber \\
& + \sin(2\alpha)\cos\beta \cos\gamma] \nonumber \\
&-Q_{3z''^2-r^2}(\theta'',\phi'') \frac{\sqrt{3}}{2} \sin^2\beta \sin\gamma  
\label{dxy}
\end{align}
we get the contribution for the $(0,0,4n+2)$ reflection as determined by the local $5d$ Re orbitals, in the frame $x''y''z''$: 

\begin{align}
Q_{xy}(\theta,\phi) & \simeq 0.245 Q_{x''^2-y''^2}(\theta'',\phi'')-0.539 Q_{3z''^2-r^2}(\theta'',\phi'')\nonumber \\
& +0.396 Q_{x''y''}(\theta'',\phi'') +0.631 Q_{y''z''}(\theta'',\phi'')\nonumber \\
& -0.307 Q_{x''z''}(\theta'',\phi'')   
\label{dxybis}
\end{align}

The first line in Eq.~(\ref{dxybis}) refers to the contribution of the $e_g$ orbitals to the reflection and the last two lines to the contribution of the $t_{2g}$ ones. We remark that the choice of $x''$, $y''$ and $z''$ orbitals in the local basis is arbitrary, so that the only physical information is the total $t_{2g}$ weight and the total $e_g$ weight. Squaring the first two coefficients and summing them up gives a $35\%$ contribution from $e_g$ orbitals to the reflection and squaring the last three coefficients and summing them up gives a $65\%$ contribution from $t_{2g}$ orbitals, as reported in Section IV.


\begin{thebibliography}{99}
\bibitem{notaSS}
We remind that the S and P channels correspond to polarization perpendicular to and within the scattering plane, respectively. In the notations SS, SP, PS, PP, the first label refers to the incoming photons, the second to the outgoing ones.
\bibitem{sergienkoJPSJ}
We used the Re labeling of the International Tables for Crystallography, No.~227, origin choice 1 for $16c$ Wyckoff positions. Compared to the notation used by Sergienko and Curnoe \cite{sergienko}, we remark that our $\vec{r}_2$ is their $\vec{r}_4$, and vice-versa.
\bibitem{noteL23}
We note that the very different behavior of the L$_2$ and L$_3$ edges found in the iridates does not occur in the present case due to the differing d electron count along with the trigonal environment of the Re sites.
\bibitem{noteOs}
In that regard, it would be helpful to have similar NMR data on the Os compound, Cd$_2$Os$_2$O$_7$, where magnetic order is seen.
\bibitem{notefin}
The two other equivalent  bonds in the cubic phase, Re$_1$-O$_3$-Re$_3$ and Re$_1$-O$_1$-Re$_4$, are practically unchanged in the tetragonal phase, with both being $139.5^\circ$.
\bibitem{harternote}
This possibility was suggested to us by John Harter in a personal communication.
\end{thebibliography}
\end{document}